\documentclass[preprint,12pt,authoryear]{elsarticle}
\usepackage{float}
\usepackage{xcolor}
\usepackage{amsmath,amssymb}
\usepackage{stmaryrd}
\usepackage{algorithm}
\usepackage[noend]{algpseudocode}
\usepackage{nomencl}
\usepackage{ulem}
\makenomenclature

\makeatletter
\def\BState{\State\hskip-\ALG@thistlm}
\makeatother

\newlength\myindent
\usepackage{physics}
\usepackage{tabularx}

\usepackage{amssymb}

\journal{International Journal of Plasticity}

\begin{document}

\begin{frontmatter}

\title{A stochastic discrete slip approach to microplasticity: Application to submicron W pillars}

\author[inst1,inst2]{Carlos J. Ruestes\corref{cor1}}
\cortext[cor1]{Corresponding author: Carlos J. Ruestes e-mail:carlos.ruestes@imdea.org}
\affiliation[inst1]{organization={IMDEA Materials Institute},
            addressline={C/Eric Kandel 2}, 
            city={Getafe},
            postcode={28906}, 
            state={Madrid},
            country={Spain}}

\affiliation[inst2]{organization={Instituto Interdisciplinario de Ciencias Básicas (ICB), Universidad Nacional de Cuyo UNCuyo-CONICET, Facultad de Ciencias Exactas y Naturales},
            addressline={Padre Contreras 1300}, 
            city={Mendoza},
            postcode={5500}, 
            state={Mendoza},
            country={Argentina}}

\author[inst3,inst1]{Javier Segurado}

\affiliation[inst3]{organization={Universidad Politécnica de Madrid, Department of Materials Science},
            addressline={E.T.S.I. Caminos}, 
            city={Madrid},
            postcode={28040}, 
            state={Madrid},
            country={Spain}}

\begin{abstract}
A stochastic discrete slip approach is proposed to model plastic deformation in submicron domains. The model is applied to the study of submicron pillar ($D~\leq~1\mu m$) compression experiments on tungsten (W), a prototypical metal for applications under extreme conditions. 
Slip events are geometrically resolved in the specimen and considered as eigenstrain fields producing a displacement jump across a slip plane. This novel method includes several aspects of utmost importance to small-scale plasticity, i.e. source truncation effects, surface nucleation effects, starvation effects, slip localization and an inherently stochastic response. Implementation on an FFT-spectral solver results in an efficient computational 3-D framework.
Simulations of submicron W pillars ($D~\leq~1\mu m$) under compression show that the method is capable of capturing salient features of sub-micron scale plasticity. These include the natural competition between pre-existing dislocations and surface nucleation of new dislocations. Our results predict distinctive flow stress power-law dependence exponents as well as a size-dependence of the strain-rate sensitivity exponent. The results are thoroughly compared with experimental literature.
\end{abstract}


\end{frontmatter}



\section{Introduction}
\label{sec:intro}

Progress towards fulfilling the promise of nanoscale solutions to engineering problems hinges 
on the development of computational and characterization techniques for an adequate assessment of the mechanical response of the resulting materials and parts. 
For instance, micropillar compression testing \citep{uchic2004sample,uchic2005methodology} can provide critical data for the development of a number of technologies at different engineering scales, from the micro/nanoscale 
(e.g micro/nano electromechanical systems -MEMS / NEMS-, microsensors and actuators and microscale energy harvesting and storage systems) to the macroscale. The technique is instrumental in establishing correlations between the microstructure and the mechanical properties of materials \citep{greer2011plasticity,dehm2018overview}. The analysis and interpretation of the experimental data obtained often relies on advanced computational simulations at different length scales, which aim at determining the underlying deformation micromechanisms responsible of the observed behavior.

At the lowest scale, molecular dynamics simulations (MD) were used to study the interaction of dislocations with pillar surfaces, leading to the proposal of a surface-induced cross-slip mechanism \citep{weinberger2008surface}. In addition, several studies have been published with different focuses, such as unraveling the tension-compression asymmetry observed experimentally \citep{healy2014molecular} or analyzing the origin of size effects observed experimentally \citep{yang2021micro,yaghoobi2016size}, to name a few. Nevertheless, the detailed nature of atomistic simulations, together with integration timesteps of the order of 1-3 fs, enforces the use of ultra-high strain rates of $10^7/s$ and above and very small pillar sizes ($D~<~100~nm$). 
The comparison between MD simulations and micropillar compression experiments is thus limited by the inherent drawbacks associated with the simulation technique. Still, MD has made fundamental contributions to the understanding of source nucleation in micropillars and nanowires \citep{weinberger2008surface,zhu2008temperature}, even allowing to estimate nucleation stresses on pristine fcc nanowires within 20\% of the experimental values \citep{jennings2013modeling}.

At the mesoscale, discrete dislocation dynamics simulations (DDD) have been used to provide insights into the mechanisms responsible for the experimentally observed staircase stress-strain behavior \citep{tang2008dislocation} as well as athermal size-dependent strengthening \citep{rao2008athermal}. The technique has also been used to study the effect of specimen size on plastic flow and work-hardening \citep{el2009role} and also to study the characteristics of source truncation controlled flow
behavior for submicron FCC single crystal micropillars \citep{cui2014theoretical}. Moreover, in order to capture surface-related effects, MD simulation-derived models of surface nucleation \citep{ryu2011entropic,zhu2008temperature} and surface-induced cross-slip \citep{weinberger2008surface} have been incorporated into DDD frameworks \citep{ryu2013plasticity,ryu2015stochastic,hu2019predicting}, allowing for the exploration of pillar sizes in the range of 100 nm to a few microns. The technique has also been used to develop stochastic models for the onset of plasticity in micro- and nano-scale structures \citep{shao2014stochastic}. In spite of their advantages, DDD simulations still suffer from timescale limitations (few ns to few $\mu$s) and typical strain rates used are in the range of $10^3/s - 10^6/s$, though strain rates as small as 0.1/s are possible depending on the computational resources \citep{fan2021strain}. Both techniques, MD and DDD, typically rely on an implementation in the form of massively parallel solvers to obtain results in a reasonable time, which in turn impose important hardware requirements. 

At the continuum scale, several groups had opted for finite element simulations (FE) with different material model approaches. FE simulations using simple isotropic plasticity have
proven useful in providing recommendations for experiments design and reliable testing \citep{zhang2006design,kiener2009micro}. Simulations using  crystal plasticity constitutive equations (CPFE) simulations have been successfully used to study the effects of initial crystallographic orientation, diameter-to-length ratio and friction on the response of Cu single crystalline pillars \citep{raabe2007effects}. CPFE has also been used for the study of polycrystalline Ni-based superalloy micropillars \citep{CRUZADO2015242}. The use of CPFE models has also extended to size effects in Al pillar compression, focusing on a continuum description of starvation \citep{JERUSALEM201293}.
CPFE relies on homogenization of the evolution of large dislocation ensembles and on scale-separation, hypotheses that can be perfectly assumed when studying the deformation of large crystal grains or specimens of several microns. However, in small-scale testing the number of dislocations in the specimen can be very small (just a few dislocations in the full specimen), such that their effect on the deformation is localized in specific planes and show an stochastic nature. These aspects cannot be homogenized in a continuum-scale law.

In order to fill gaps between DDD and CPFE models several attempts have been made to extend standard CPFE models to account for some characteristic features of small-size tests (i.e. the  stochastic nature of the response and slip-band localization). Regarding the stochastic response, \cite{ng2008stochastic} proposed a Monte Carlo model based on the survival probability and the burst size vs. stress distributions but with no further connection with the material structure. Later on,  \cite{konstantinidis2014capturing} proposed a cellular automaton, based on the gradient plasticity framework of \cite{zhang2011interpreting}, to model stochastic effects in pillar compression. In addition, \cite{lin2015stochastic} presented a crystal plasticity model in which stochasticity was introduced by defining the plastic strain as composed of a series of strain bursts whose sizes follow a power-law distribution function and whose rates are determined by a constitutive equation. More recently, a general stochastic CP formulation based on kinetic Monte Carlo, was proposed in which the active slip systems were selected based on the strain rates dictated by a mobility law depending on resolved stress and temperature \citep{yu2021stochastic}. However, the homogenized nature of the formulation prevents its direct use in applications where slip localization is expected.

The slip-band localization observed in experiments, resulting from the activation of discrete slip events in a particular plane, cannot be naturally captured by CPFE models. Some attempts have been made to emulate this localization in CP frameworks.  \cite{lin2016numerical} developed a dislocation-based crystal plasticity model in which, in order to trigger localization in a single slip band, strain-softening was added to the slip resistance evolution law and a void was introduced in the center of the pillar to trigger localization. More recently, \cite{wijnen2021discrete} presented a CPFE model adapted for simulating heterogeneous plastic deformation in single crystals. In this model, only some regions of the domain under study are modeled using CP keeping the others as elastic material. These slip areas are sampled from a probability distribution based on characteristics of the dislocation sources and plastic localization is accomplished by considering different slip bands and the weakest link principle \citep{norfleet2008dislocation}. These extended CPFE models allow to describe the strain localization in pillars but maintain a deterministic nature and do not explicitly consider plastic events by individual dislocations.

Summarizing, the mechanical response of metals and alloys under micro compression testing is typically characterized by certain features: i) samples usually have a low dislocation density and exhibit source truncation hardening;
ii) Plastic deformation is usually localized and in the form of slip bands; iii) For diameters below 150 nm, or starved samples such as those produced by mechanical annealing, surface nucleation of dislocations is likely to occur. As a consequence, plasticity takes place by events of discrete nature and the mechanical response of such structures is typically stochastic. Despite all the aforementioned computational advances, capturing all these features into a single and efficient computationally-cheap 3-D framework is still a challenge. 

The purpose of this work is two fold. Firstly, we present a multiscale computational framework capable of capturing the fundamental aspects of the mechanical response of submicron metallic parts. In this model, the displacement jumps along a cross sectional plane of the sample produced by the slip of a dislocation are introduced as eigenstrains around that plane using the Eshelby formalism. The framework also accounts for surface nucleation by implementing a model derived from MD simulations. The stochastic nature of the mechanical response is captured by means of a kinetic Monte Carlo selection process. In this kMC approach, the events correspond to the single slip events produced by dislocations, whose probabilities are dictated by physically-based laws. This approach is implemented in an FFT-based solver resulting in a very efficient and computationally-cheap framework that allows to model complex 3D geometries.
Secondly, we revisit micropillar compression experiments of single crystalline bcc submicron pillars ($D~\leq~1\mu m$) with focus on tungsten (W). 
We show that not only our method can successfully capture fundamental aspects of the mechanical response at these scales, but also allows for detailed investigation of size-effects, strain-rate effects and statistics of strain-bursts.  

\section{Methodology}
\label{sec:methods}

The framework developed assumes dislocation-mediated plasticity as the underlying inelastic deformation micromechanism. We consider pre-existing dislocations as well as newly developed dislocations due to surface nucleation.
The amount of pre-existing dislocations $N_{d}$ on a pillar of diameter $D = 2 R$, height $H$, and initial dislocation density $\rho$, is determined as
\begin{equation}
N_{d} = int(\rho~ \frac{\pi~R^2~H}{L_{d}})
\end{equation}

\nomenclature{\(\rho\)}{Dislocation density}
\nomenclature{\(R\)}{Pillar radius}
\nomenclature{\(D\)}{Pillar diameter}
\nomenclature{\(H\)}{Pillar height}
\nomenclature{\(L_d\)}{Average length of a dislocation segment}
\nomenclature{\(N_{SN}\)}{Number of potential surface nucleation sites}

where $L_d$ is the average length of a dislocation segment, here taken as R for simplicity. 
Then, $N_d$ slip planes are randomly selected along the pillar axis with normal directions consistent with the available slip systems. 

Surface nucleation can potentially happen at any position on the pillar surface. From a practical point of view, in order to consider a dislocation nucleation and its subsequent slip in our framework, a large but finite number of nucleation sites $N_{SN}$ is considered. In practice, $100 < N_{SN} < 200$. The positions of the resulting slip planes are randomly selected along the pillar axis, and their orientations are also based on the available slip systems. 

In our framework, dislocations are not explicitly considered, but the effect of their extension when reaching the surface of the pillar. This effect consists in a relative displacement between the upper and lower parts of their plane and experimentally would define a slip trace. This description is equivalent to the 
definition of a closed dislocation loop in phase-field dislocation dynamics as an inclusion with an eigenstrain (an Eshelby inclusion) \citep{wang2001nanoscale,rodney2003phase}. In our case, only the final stage of the loop is considered which consists then in a loop with a shape defined by the intersection of the pillar with the slip plane; see Sec. \ref{kinematics} for details. The different parts of the model are introduced in the next sections.

\subsection{Kinematics of a slip event}
\label{kinematics}

\begin{figure}[h]
    \centering
    \includegraphics[width=0.8\textwidth]{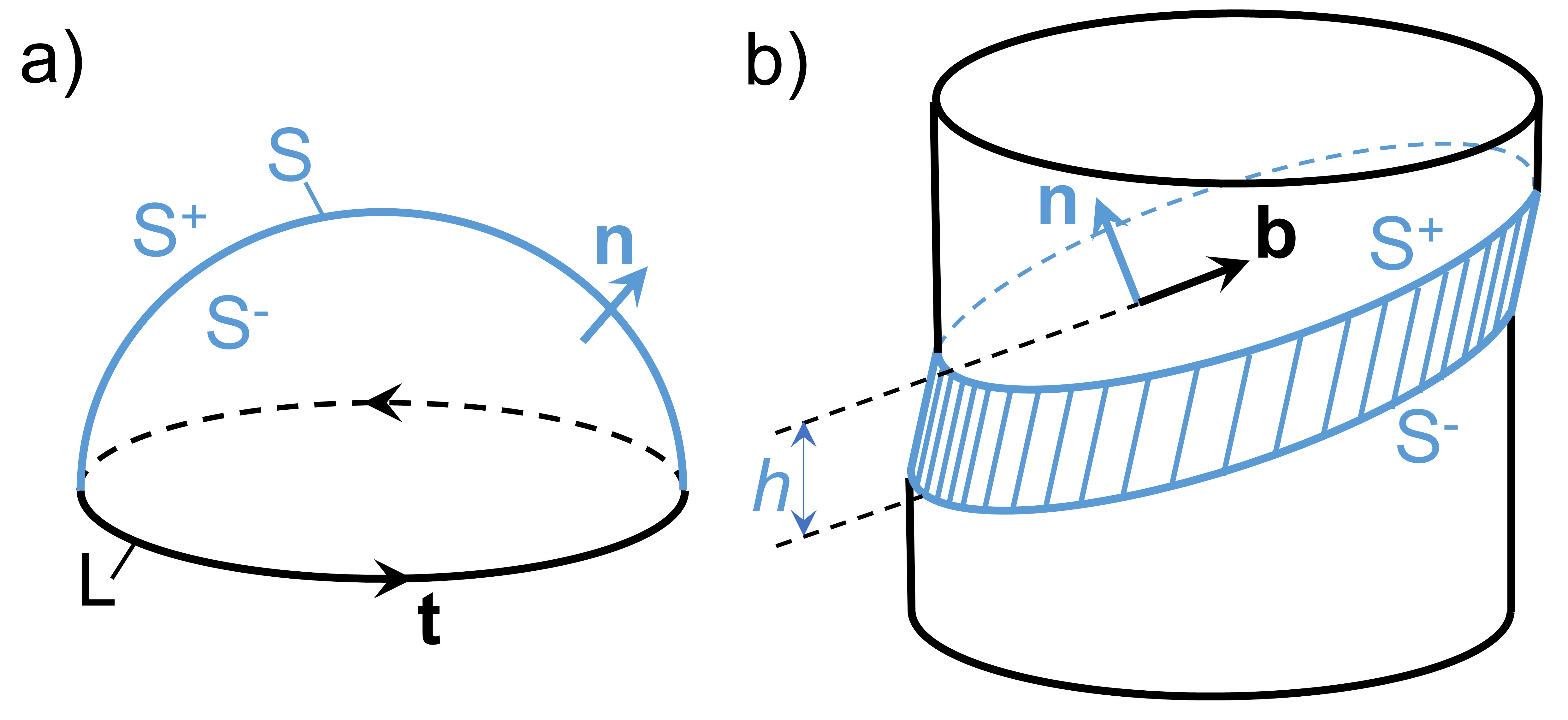}
    \caption{Dislocation eigenstrain formalism and the application to a single crystalline pillar. (a) A dislocation loop $L$ acting as a boundary of a dislocation surface $S$. By sweeping the upper part $S^+$ with respect to $S^-$ in an amount $b$, a dislocation is effectively introduced. (b) For a dislocation sweeping a pillar, the dislocation can be replaced by a plate-like Eshelbian inclusion of thickness h.  }
    \label{fig:eigenstrainformalism}
\end{figure}

First we derive the elastic displacement introduced by a dislocation loop considering it as an equivalent Eshelbian inclusion. Under this representation, the loop is idealized by moving the upper side of surface $S$, denoted by $S^+$, by $b$ with respect to the lower side $S^-$, as shown in Fig. \ref{fig:eigenstrainformalism}.a. Let $\mathbf{n}$ be the normal to the loop, then the relative displacement between the upper and lower parts of the loop in direction $\mathbf{n}$ is given by 
\begin{equation}
    \llbracket \mathbf{u} \rrbracket_n (\mathbf{x})=\mathbf{b} \ \delta_S(\mathbf{x}), 
\end{equation}
where $\mathbf{b}$ is the burgers vector of the loop and $\delta_S(\mathbf{x})$ denotes the surface Dirac delta function, that is infinite if $\mathbf{x} \in S$ and zero elsewhere. If the loop is fully embedded in a material, the resulting elastic fields of this incompatible deformation can be obtained using Green's function \citep{mura2013micromechanics}. In our model, the only stage considered is when the loop has filled all the pillar cross section.

The displacement jump due to a full plane slip has been considered in the past using a strong discontinuity approach in the context of finite elements and discrete dislocation dynamics \citep{Romero2008} . However, introducing this discontinuity in other numerical frameworks as FFT imply smoothing out the jump to make
 the model numerically tractable. This can be done by distributing the jump along a thin band of size $h$, such that it renders the same total displacement as in the initial ideal case. In a simplified way, the line integral along a path parallel to $\mathbf{n}$ corresponds to
\begin{equation}
    \int_{z=-h/2}^{z=h/2} \boldsymbol{\varepsilon} (\mathbf{x}) \mathrm{d}z = \llbracket \mathbf{u} \rrbracket_n 
\end{equation}

Extending the use of the equivalent Eshelbian inclusion, Figure \ref{fig:eigenstrainformalism}.b depicts a scenario in which the cross-section of cylindrical pillar has been fully swept in an amount of $b$ by a single dislocation loop. The associated displacement is linearly distributed over a thickness $h$ in the direction perpendicular to the loop. 
As explained later on (Sec. \ref{sec:FFTmethod}), the simulation domain is discretized
in a box of [Nx, Ny, Nz] voxels, so in practice, $h$ is taken as twice the distance between voxels \citep{capolungo2019gd}. As presented in the Supplementary Material file, the method is relatively insensitive to the election of the voxel size. 
In consequence, the strain $\boldsymbol{\varepsilon}^{EIG~(i)}(\mathbf{x})$ associated with a slip event $i$, is taken as: 

\begin{equation}
    \label{eigenstrain}
    \boldsymbol{\varepsilon}^{EIG~(i)}(\mathbf{x}) =
    \begin{cases}
        \frac{b}{h}~(\boldsymbol{s}^i \otimes  \boldsymbol{n}^i), ~ if ~\mathbf{x} ~\epsilon ~plastic~region \\
        0,~elsewhere
    \end{cases}
\end{equation}

\nomenclature{\(\boldsymbol{\varepsilon}^{EIG~(i)}\)}{Eigenstrain corresponding to slip event i}
\nomenclature{\(b\)}{Burgers vector modulus}
\nomenclature{\(h\)}{thickness of plastic region slip band}
\nomenclature{\(M\)}{Total number of slip events}
where $b$ is the burgers vector modulus, $h$ is the thickness of the slip band and $\mathbf{s}^i$ and $\mathbf{n}^i$ stand for the burgers vector direction (slip) and normal direction to the slip plane, $\mathbf{s}^i\otimes\mathbf{n}^i$ being the Schmid tensor. Consequently, for a pillar that underwent $M$ slip events, the total eigenstrain field $\boldsymbol{\varepsilon}^{EIG}(\mathbf{x})$ is the superposition of all the eigenstrains applied, 

\begin{equation}
    \label{sumEig}
    \boldsymbol{\varepsilon}^{EIG}(\mathbf{x}) = \sum_{i=1}^{M} \boldsymbol{\varepsilon}^{EIG~(i)}(\mathbf{x})
\end{equation}

\nomenclature{\(\boldsymbol{\varepsilon}^{EIG}\)}{Total eigenstrain field}

\subsection{Constitutive laws for the slip events}
\label{physec}

The laws linking the stress at a given plane with its slip rate will be presented in next sections.
\subsubsection{Pre-existing dislocations}
\label{preexisting}
The model assumes that the pillar contains a certain number of dislocations which will drive the slip events. Two types of pre-existing dislocations are considered, these are single-arm sources and screw dislocation segments. The former aid in the introduction of source truncation effects, while the latter allow for source exhaustion effects.

\paragraph{\textit{Single-Arm Sources}}

A dislocation with a pinning point in the interior of the crystal and a free end on the surface acts as a source of dislocations in small pillars and is commonly termed as Single-Arm Source (SAS) \citep{parthasarathy2007contribution}. In the present model, these sources are randomly distributed in the pillar volume, assigning to each source a particular slip plane. The critical resolved stress for activating a SAS is determined by its arm length $\lambda$,
\begin{equation}
    \label{SAS_CRSS}
    \tau^{SAS}_{CRSS} = \frac{\alpha_{SAS}~\mu~b}{\lambda}+\tau_0 
\end{equation}
\nomenclature{\(\tau^{SAS}_{CRSS}\)}{Critical resolved shear stress to activate a Single Arm Source}
\nomenclature{\(\beta\)}{geometrical constant for SAS model}
\nomenclature{\(\mu\)}{Shear modulus}
\nomenclature{\(\tau_0\)}{Friction stress}
\nomenclature{\(\lambda\)}{Single Arm Source length}

where $\alpha_{SAS}$ is a source-strength coefficient that considers the nature of the source, its length and material, here taken as 0.6 after \citep{rao2007estimating}. $\mu$ is the shear modulus, $b$ the Burgers vector and $\tau_0$ the friction stress. 
In their presentation of the SAS model, \cite{parthasarathy2007contribution} include a Taylor-type term in eq. \ref{SAS_CRSS} to account for dislocation forest interactions. 
This term is not considered here since, for D $\leq$ 1 $\mu$m specimens and typical dislocation densities of the order of $10^{12} m^{-2} - 10^{13} m^{-2}$, the number of dislocations present in the pillar becomes too small to homogenize their effect ($N_d ~\leq~ 10)$.  
Under such conditions, weak interactions are expected. Thus, we essentially neglect hardening. For pillar diameters above 1 micron, hardening should be incorporated, as explained in Sec. \ref{discussion}.

In order to generate a random arrangement of SAS, for each source a random pining point is generated in the pillar interior. Then, a slip plane among the characteristic active planes of the metal lattice is randomly assigned. The source is defined as a straight line of length $\lambda$ connecting the pinning point with the nearest point in the elliptical boundary resulting from the intersection of the plane with the pillar surface. For more details on this model, the reader is referred to \citep{parthasarathy2007contribution}. 

This model has been widely used on experimental studies \citep{abad2016temperature}, DDD simulations \citep{hu2019predicting}, Crystal Plasticity simulations \citep{gu2021statistical}, as well as on recent discrete slip plane models \citep{wijnen2021discrete}. In the present work, it is assumed that single-arm sources remain immutable after  activation. Changes in SAS could be potentially included in the model, for example the incorporation of SAS destruction as proposed in the work by \cite{cui2015theoretical} on coated submicron pillars.

\paragraph{\textit{Screw dislocation segments}}

In addition to SASs, we also consider the possibility of pre-existing pure-screw segments. In essence, such dislocations are not pinned and can sweep across the slip plane, exiting the cylinder as they reach its surface. Under these operating conditions, screw segments allow to capture the so-called "dislocation-starvation" behavior \citep{greer2006nanoscale}. In this case, the resolved stress necessary to move the dislocation is only the lattice friction,

\begin{equation}
    \tau^{screw}_{CRSS} = \tau_0
\end{equation}

\nomenclature{\(\tau^{screw}_{CRSS}\)}{Critical resolved shear stress to activate a screw dislocation}

The position of the screw dislocation within the pillar cross-section is randomly chosen. As shown in section \ref{sec:mechanicalannealing}, this allows capturing different cases of mechanical annealing observed in the literature, with variable stress drop due to the activation of these sources. It is worth noting that in our implementation, dislocation cross-slip due to image forces in the vicinity of the free surfaces is not considered. This is further discussed in Sec. \ref{discussion}.

\subsubsection{Surface nucleation model}
\label{surfacemodel}

For pillars and nanowires in the range of a few tenth of nanometers to a few hundred of nanometers, virtually defect-free microstructures are not unlikely. 
Under these conditions, mechanical testing of nanopillars and nanowires reveal extremely high tensile stress and atomistic simulations point to surface nucleation of dislocations as the source of plasticity \citep{zhu2008temperature}. In consequence, the framework developed includes surface nucleation effects.

The surface nucleation model adopted here was initially proposed after atomistic simulations \citep{zhu2008temperature,ryu2011entropic} and later adopted on discrete dislocation dynamics (DDD) simulations \citep{ryu2015stochastic,hu2017strain,hu2019predicting}. Briefly, within a time span $\Delta t$, the probability of a surface nucleation event in a pillar of surface $S$ is given by
\begin{equation} \label{probability}
P = \nu_0 ~ exp [-\frac{F(\sigma,T)}{k_B ~ T}] \cdot \frac{S}{b^2} \cdot \Delta t.
\end{equation}
\nomenclature{\(k_B\)}{Boltzmann constant}
\nomenclature{\(\nu_0\)}{Attempt frequency for Surface Nucleation model}
\nomenclature{\(S\)}{Pillar surface area}
\nomenclature{\(F(\sigma,T)\)}{Activation free energy for Surface Nucleation Model}
Here, $\nu_0$ is an attempt frequency and $k_B$ the Boltzmann constant.
Thus, the ratio $\frac{S}{b^2}$ stands for the number of potential nucleation sites. Finally, $F(\sigma,T)$ is an activation free energy that depends on stress and temperature through

\begin{equation}
F(\sigma,T) = (1-\frac{T}{T_d}) \cdot F_0(\sigma)
\end{equation}

\nomenclature{\(T_d\)}{Disorder Temperature for SN model}

with

\begin{equation}
F_0(\sigma) = A \cdot (1 - \frac{\sigma}{\sigma_{athm}})^{\alpha_{SN}}.
\end{equation}

\nomenclature{\(A\)}{zero-stress, zero-temperature activation energy for SN model}
\nomenclature{\(\alpha_{SN}\)}{Constant for stress dependency of SN model}
\nomenclature{\(\sigma_{athm}\)}{Athermal nucleation stress for SN model}

Here, $A$ is the zero-stress, zero-temperature activation energy, while $\alpha_{SN}$ and $T_d$ are constants corresponding to the stress and temperature dependencies, respectively. $T_d$ is usually taken as half the bulk melting temperature \citep{zhu2008temperature}. Finally, $\sigma_{athm}$ is an athermal nucleation stress, corresponding to the tensile stress required for surface nucleation at zero-temperature. The possibility of surface nucleation is considered once $P=1$. 

The parameters $A$, $T_d$, $\alpha_{SN}$ and $\sigma_{athm}$ are material-dependent and their calibration require dedicated studies, either based on atomistic simulations \citep{zhu2008temperature} or on experiments \citep{chen2015measuring}. To the best of the authors' knowledge, there are no such studies on W. Therefore, we opt to fit these parameters to the available experimental data, following suggestions by \cite{chen2015measuring}. Nanopillar compression and nanowire tensile test for [100]-oriented W are extremely scarce in the literature. \cite{srivastava2021influence} report yield stresses of the order of 3.5 GPa in their 100 nm diameter pillar compression tests, whereas \cite{cordoba2017suspended} report yield stresses of 6.6 $\pm$ 1.2 GPa for their 93 nm suspended wires under bending. Therefore, we opted to fit the parameters of eqn. \ref{probability} to a 100 nm target yield strength of 4.5 GPa, in between the mean value of  \cite{srivastava2021influence} and the lower bound of the range reported by \cite{cordoba2017suspended}. The resulting parameters are included in Table \ref{table:SN1} after the fitting procedure presented in the Supplementary Material file. 
The surface nucleated dislocations are considered as pure screw segments, as described in  the previous subsection.

\begin{table}[]
\begin{tabular}{llll}
\hline
Symbol                      & Property                                             & Value & Unit \\ \hline
$T_d$                        & Disorder temperature                                 & 1800  & K    \\
A                           & zero $\sigma$, zero $T$ activation energy & 6     & eV   \\
$\alpha_{SN}$   & Fitting exponent stress dependence                   & 4     & -    \\
$\sigma_{athm}$ & Athermal nucleation stress                           & 16    & GPa  \\ \hline
                            &                                                      &       &     
\end{tabular}
\caption{Surface nucleation model parameters.}
\label{table:SN1}
\end{table}

\subsubsection{Dislocation mobility and displacement rates}
\label{mobility_and_rates}

The  displacement rate of the pillar in the slip direction $\mathbf{s}^i$  due to the slip produced in a plane $i$ by the movement of a dislocation can be obtained using a reasoning similar to Orowan's equation,
\begin{equation}
\dot{\mathbf{u}}^i_p = b^i \frac{v^i}{D^i}
\mathbf{s}^i \label{eq:u_p1}
\end{equation}

\nomenclature{\(M^{i}\)}{Schmid factor of slip system associated to source i}
\nomenclature{\(v^{i}\)}{Velocity of the dislocation associated to source i}
where $b^{i}$ and $v^{i}$ are the Burgers vector modulus,  and dislocation velocity 
corresponding to source $i$, respectively. $D^i$ corresponds to the length in direction $\mathbf{s}^i$ swept by a single dislocation when fully crossing the slip plane $i$.
If the unit vector $\mathbf{e}$ stands for the pillar axis and $D$ for its diameter, then $D^i=D/(\mathbf{n}^i\cdot\mathbf{e}$), and the vertical displacement rate corresponds to
\begin{equation}
\dot{u}^i_p=\dot{\mathbf{u}}^i_p\cdot\mathbf{e} = b^i \frac{v^i}{D/\mathbf{n}^i\cdot\mathbf{e}}
\mathbf{s}^i\cdot\mathbf{e}= \frac{b^i \ v^i M^i}{D}
\label{eq:u_p2}
\end{equation}
\nomenclature{\(\dot{u}^i_p\)}{Vertical displacement rate on a plane i}
where $M^i$ is the Schmid factor of slip system associated to source $i$ projected in direction $\mathbf{e}\otimes\mathbf{e}$.

In order to compute the displacement rate for each plane $i$, the associated dislocation velocity $v^{i}$ must be determined. 
Dislocation-mediated plasticity in bcc metals can take place in 48 bcc slip systems including the $\{110\}$,$\{112\}$ and $\{123\}$ families of slip planes. Interestingly, some works propose the decomposition of $\{112\}$ and $\{123\}$ slip planes on alternating $\{110\}$ slip planes  \citep{christian1983some,marichal2013110,ruestes2014plastic}. Based on this, and for the sake of simplicity, we consider bcc metals with only $\{110\}$ slip planes ($\bar{b}=1/2~a_0~<111>$ dislocations). That is, $\vert \bar{b} \vert$ is unique and equal to $a_0 ~ \sqrt{3}/2$, with $a_0$ the bcc lattice parameter. 


The mobility laws for bcc tungsten dislocations have been deeply studied using atomistic simulations 
\citep{cereceda2013assessment,stukowski2015thermally,cereceda2016unraveling,po2016phenomenological}. For tungsten, and considering relatively low temperatures, dislocation velocities can be estimated 
by: 

\begin{equation}
\label{v_alfa}
v = 
    \begin{cases}
        \frac{\mu_0 ~h'~(\zeta - w)}{b} exp (-\frac{\Delta H(\tau_{eff})}{k_B~T} ), &\Delta H > 0\\
        \frac{\tau_{eff} ~ b}{B}, &\Delta H \leq 0
    \end{cases}
\end{equation}

\begin{equation}
\label{delta_H}
\Delta H(\tau_{eff}) = \Delta H_0 \lceil 1 - (\frac{\tau_{eff}}{\tau_*})^p   \rceil ^q 
\end{equation}
\nomenclature{\(\mu_0\)}{Attempt frequency for the dislocation mobility law}
\nomenclature{\(h'\)}{distance between Peierls valleys}
\nomenclature{\(\zeta\)}{mean dislocation segment length}
\nomenclature{\(w\)}{kink-pair width}
\nomenclature{\(p\)}{energy profile parameter}
\nomenclature{\(q\)}{energy profile parameter}
\nomenclature{\(\Delta H_0\)}{kink-pair formation energy}

\begin{table}[]
\begin{tabular}{llll}
\hline
Symbol     & Property                         & Value                      & Unit \\ \hline
$\mu_0$    & Attempt frequency                & 9.1 $10^{11}$ & $s^{-1}$   \\
$h'$         & Distance between Peierls valleys & $\sqrt{6}/3$                  & $a_0$   \\
$w$          & Kink pair width                  & 11                         & b   \\
$\zeta$    & Mean dislocation segment width   & 25                         & b   \\
$\tau_P$ & Peierls stress                   & 2.03                       & GPa  \\
$\Delta H_0$    & Kink-pair energy at 0 K                          & 1.63                       & eV   \\
$p$          & Energy profile parameter                                 & 0.86                       & -    \\
$q$          & Energy profile parameter                                 & 1.69                       & -    \\
$B$          & Phonon drag coefficient          & 9.8 $10^{-4}$                 & Pa*s \\ \hline
\end{tabular}
\caption{Parameters for the dislocation mobility law. Obtained from molecular dynamics simulations \citep{cereceda2013assessment,stukowski2015thermally,cereceda2016unraveling,po2016phenomenological}}
\label{table:dislocmobility}
\end{table}

In the expression,  $\mu_0$ is an attempt frequency, $h'$ is the distance between Peierls valleys, $\zeta$ is the mean dislocation segment length for the slip system, $w$ is the kink-pair width and $\Delta{H_0}$ is the kink-pair formation energy. Parameters $p$ and $q$ allow for the adjustment of the "tail" ($\tau_{eff}\simeq0$) and "top" ($\tau_{eff}\simeq\tau_*$) of the energy profile $\Delta H$, respectively \citep{kocks1975thermodynamics}. The parameters for W are presented in Table \ref{table:dislocmobility}.

The upper part of the definition ($\Delta H > 0$) corresponds to the kink-pair thermally-activated regime, typical of bcc metals \citep{kocks1975thermodynamics}. The lower part, in turn, corresponds to the phonon-drag-dominated regime, typical of high stress / high strain-rate conditions \citep{po2016phenomenological}. In Eq. \eqref{delta_H} $\tau_{eff}$ represents the effective resolved stress on the plane $i$,
\begin{equation}
\tau_{eff} = \vert \overline{\tau}^{i} \vert - \tau_a ^ {i} \label{eq:taueff}
\end{equation}
where $\overline{\tau}^{i}$ is the resolved shear stress, averaged over the plane, for source $i$ resulting from the Cauchy stress projected in the particular system to which that source belongs to,
\begin{equation}
\label{tau_i}
  \tau^{i} = \boldsymbol{\sigma} : (\mathbf{s}^{i} \otimes \mathbf{n}^i).
\end{equation}
In Eq. \eqref{eq:taueff} $\tau_a$ is the resistance to slip due to athermal barriers, as dislocation groups, precipitates, which  are not considered in the present study 
($\tau_a ^ {i} = 0$). Finally, $\tau_*$ represents the part of the resistance to slip due to thermally activatable obstacles. In the case of pure W the only thermal barrier is the Peierls resistance, hence $\tau_* = \tau_P$.


In essence, the use of eq. \ref{v_alfa} for the calculation of the pillar displacement rates (eq. \ref{eq:u_p2}) allows to incorporate the thermally-activated motion character of screw dislocations at low stresses and temperatures, as well as the phonon-drag-dominated regime for newly surface-nucleated dislocations under high stresses. Equations \ref{eq:u_p2}-\ref{v_alfa} allow for the effect of each plane $i$ in the displacement rate of the pillar.

\subsection{Kinetic Monte Carlo procedure for stochastic slip events}
\label{kMC_details}


In our model, the activation of the previously described microscopic slip mechanisms is controlled by a kinetic Monte Carlo selection process. 
Considering a displacement-controlled micropillar compression experiment with an applied strain rate of $\dot{\varepsilon}_0$, the velocity of the upper part of the pillar corresponds to $\dot{u}_0= H\dot{\varepsilon}_0$. This displacement is accommodated by either elastic or plastic deformation. The plastic contribution to the displacement of the pillar corresponds to the superposition of the displacement produced by  slip in the active planes
$$ 
\dot{u}_p= \sum_{i} \dot{u}^i_p
$$
with $\dot{u}^i_p$ given by eq. \ref{eq:u_p2}. The elastic displacement rate then corresponds to 

\begin{equation}
\label{eqn.dotu}
    \dot{u}_E=\dot{u}_0-\dot{u}_p.
\end{equation}

\cite{yu2021stochastic} proposed a rejection-free kinetic Monte Carlo algorithm to obtain the stress-strain evolution in a deforming single crystal in which the strain rates of the different slip systems provide the set of event rates. In this work, this idea is generalized to a full deforming specimen solving the macroscopic stress-strain response as well as the spatial distribution of the fields involved. At time $t^n$, each of the $i$ escape pathways has a rate constant $\dot{u}_p^i$, that characterizes the probability per unit time that the system escapes to that state $i$. These rates can be used to assemble an array of partial sums representing the accumulated rate of all the objects up to and including object j,
\begin{equation}
    \label{accumulatedrates}
    r_j = \sum_{i=1}^{j} \dot{u}_p^{i} + r_0
\end{equation}
with $r_0=\dot{\varepsilon_0}$ 
Then, the total escape rate is the sum of all the rates

\begin{equation}
    \label{total_rate}
    r_{tot} = \sum_{i=1}^{N} \dot{u}_p^{i} + r_0
\end{equation}
$N$ being the total number of sliding planes considered. In agreement with rejection-free kMC theory, the next event to be executed is the $k^{th}$ process satisfying
\begin{equation}
    r_{k-1} < \xi_1 r_{tot} \leq r_{k}
\end{equation}

\begin{figure}[h]
    \centering
    \includegraphics[width=0.8\textwidth]{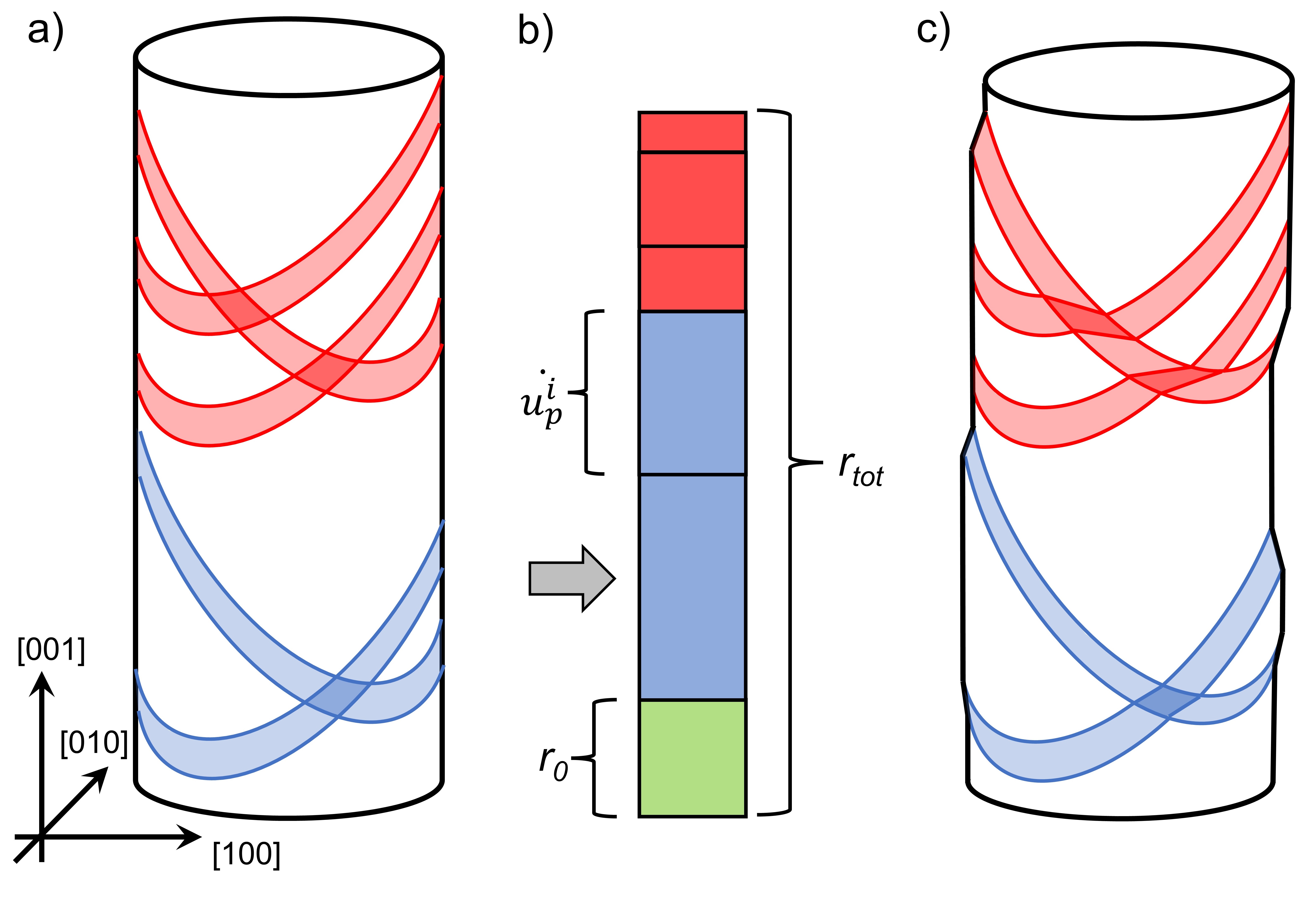}
    \caption{Schematic of stochastic selection process. a) a number of plastic regions are randomly selected both for pre-existing dislocations (blue) and possible surface nucleation sites (red). b) Upon loading and based on eqns. \ref{eq:u_p2} through \ref{total_rate}, a sampling array of events in built. The total rate $r_{tot}$ is composed of the prescribed applied strain rate $r_0$ and the vertical displacement rates of each plastic region. In general, the value of each vertical displacement rate $\dot{u}^i_p$ is different from the rest of the displacement rates during each time step, as represented by the different box sizes. c) After selection of a number of events using rejection-free kMC, the final deformed state of the pillar reflects the operation of pre-existing dislocations as well as a surface nucleation site. Arrow indicates most probable event. }
    \label{fig:MonteCarlo}
\end{figure}

If the event selected corresponds to the rate $r_0$, the next event will correspond to an elastic event, i.e. no inelastic deformation will be introduced in the plastic layers. Otherwise, a plastic event is chosen, for which the corresponding eigenstrain will be applied. The pillar geometry, array of displacement rates, and the resulting pillar shape after the whole process are exemplified in Figure \ref{fig:MonteCarlo}.

Then the next time step is computed as:
\begin{equation}
    \label{dt}
    \delta t^{n+1} = -\frac{log \xi_2~\Delta \varepsilon^* }{r_{tot}}
\end{equation}
such that the time is advanced as:
\begin{equation}
    \label{totalt}
    t^{n+1} = t^n + \delta t^{n+1}
\end{equation}
$\xi_1$ and $\xi_2$ are random numbers uniformly distributed in $(0,1]$ and $\Delta\varepsilon^*$ is a normalization factor ($\Delta\varepsilon^* <= 1$) to adjust the time step within physical time scale bounds \citep{yu2021stochastic}.  Once an event is executed, the strain fields and stress fields are updated. 

\paragraph{\textit{Time scale bounds}}

The absolute time scale emanating from the sampling of eq. \ref{eqn.dotu} using eqns. \ref{accumulatedrates} through \ref{dt} represents the maximum time step compatible with the rate equation under consideration. As shown by \cite{yu2021stochastic}, it is indeed possible to use an arbitrarily smaller time step without invalidating the method. A physically-reasonable upper bound for the time step is that of the total time for a single dislocation to sweep the pillar section:

\begin{equation} \label{t_i}
t^{i} = \frac{D_i}{v^{i}}
\end{equation}


In addition, the existence of $t^{i}$ also poses a constraint. On each plastic event, the eigenstrain applied cannot be larger than that of eq. \ref{eigenstrain}, which corresponds to a dislocation fully sweeping a pillar on a time $t^{i}$. Thus, $t^{i}$ is an upper bound for the time increment $\delta t^{n+1}$. As consequence, $\delta t^{n+1}/t^{i} ~ \leq 1$.
In practice, the eigenstrain applied once a plastic event is selected is: 
\begin{equation}
    \label{eigenstrain2}
    \boldsymbol{\varepsilon}^{EIG~(i)}(\mathbf{x}) =
    \begin{cases}
        (\frac{\delta t^{n+1}}{t^{i}})(\frac{b}{h})~(\hat{s} \otimes  \hat{n}), ~ if ~\mathbf{x} ~\epsilon ~plastic~region \\
        0,~elsewhere
    \end{cases}
\end{equation}

\paragraph{\textit{Weakest link principle and deformation localization}}

Under the weakest link principle, plastic flow occurs when the stress is high enough to activate the weakest source available in the crystal \citep{norfleet2008dislocation}. In our model, weaker sources with a lower activation stress will tend to produce higher dislocation velocities (eq. \ref{v_alfa}) and thus render higher plastic displacement rates (eq. \ref{eq:u_p2}). In consequence, a sample with several SASs, out of which one is significantly longer than the others, will result in a higher probability of activation of such source, thus favoring localization in the corresponding plastic region. In contrast, for a sample with several SASs of commensurate length, the kMC selection process will translate into the sequential activation of several SASs, favoring delocalization of deformation into several plastic regions. These aspects are further discussed in Sec. \ref{discussion} after the results presented in Sec. \ref{sec_resultsGeneral}. 
\subsection{Solving the mechanical problem with an FFT algorithm}
\label{sec:FFTmethod}
In order to obtain the strain and stress distribution in the pillar, which drives the activation of slip in the plastic regions, an elastic problem with eigenstrains has to be solved at each time step. An FFT-based approach will be used for this purpose.

The total strain is the sum of the eigenstrains (input) and the elastic strain $\boldsymbol{\varepsilon^e}$, which appears to enforce total strain compatibility and stress equilibrium,
\begin{equation}
    \boldsymbol{\varepsilon} = \boldsymbol{\varepsilon^e}+\boldsymbol{\varepsilon}^{EIG}.
\end{equation}
\nomenclature{\(\boldsymbol{\varepsilon}\)}{Strain field}
\nomenclature{\(\boldsymbol{\varepsilon^e}\)}{Elastic strain field}
Under linear elasticity, stress is related to strain through
\begin{equation}
    \label{sigmaLE}
    \boldsymbol{\sigma} = \mathbb{C} : \boldsymbol{\varepsilon^e} = \mathbb{C} : (\boldsymbol{\varepsilon}-\boldsymbol{\varepsilon}^{EIG})
\end{equation}
where $\mathbb{C}$ is the stiffness tensor of the crystal. 
\nomenclature{\(\mathbb{C}\)}{Stiffness tensor}
\nomenclature{\(\boldsymbol{\sigma}\)}{Stress field}

The mechanical equilibrium corresponds to
%
\begin{equation}
    \label{divsigma_exp}
    \nabla \cdot  [\mathbb{C} : (\boldsymbol{\varepsilon}-\boldsymbol{\varepsilon}^{EIG}) ]        = \boldsymbol{0}
\end{equation}
and rearranging terms, the result is a linear differential equation 
\begin{equation}
    \label{divsigma_eq}
    \nabla \cdot  [\mathbb{C} : \boldsymbol{\varepsilon}] =  \\
    - \nabla \cdot  [\mathbb{C} : \boldsymbol{\varepsilon}^{EIG}].
\end{equation}
If the right-hand side of Eq. \eqref{divsigma_eq} is known, as in our case, the total strain $\boldsymbol{\varepsilon}$ can be directly obtained by the convolution of this term with the corresponding Green's function derivative. Moreover, if the medium is homogeneous and periodic, closed expressions for Green's function exist in Fourier space, and the solution is just a multiplication.

However, when considering the present problem in a FFT framework, the actual geometry has to be embedded in a cuboidal periodic domain $\Omega$ which also contains the outer free space. In this case, the domain can be considered as an heterogeneous microstructure formed by two phases and this microstructure can be described by 

\begin{equation}
\label{C}
\mathbb{C} = 
    \begin{cases}
        \mathbb{C}_1,~if~ \mathbf{x}~ in~ \Omega_1   \\
        \mathbb{C}_2,~if~ \mathbf{x}~ in~ \Omega_2
    \end{cases}
\end{equation}

In the simulation domain, the metallic pillar occupies the region $(\Omega_1)$ and is surrounded by an infinite compliant medium which occupies the region $(\Omega_2)$. For the solution of Eq. \ref{divsigma_eq}, considering Eq. \ref{C}, we follow the Fourier-Galerkin approach. The details of the method can be found in \citep{vondvrejc2014fft,zeman2017finite} and here only the final equations are recalled.
The problem is discretized in a box of $[N_x,N_y,N_z]$ voxels which correspond to the same number of frequencies in the discrete Fourier space. The value of $\mathbb{C}$, strain and stress fields are approximated by trigonometrical polynomial, whose coefficients are given by the FFT of the value of these fields at the voxels. Let $\mathcal{G}$ be the projection operator that, by convolution, extracts the compatible part of a tensor field \citep{vondvrejc2014fft}. The mechanical equilibrium of the discrete stress field translates into, 
\begin{equation}
    \mathcal{G} \ast \boldsymbol{\sigma} = \boldsymbol{0}
\end{equation}
where $\ast$ indicates a convolution. Introducing the value of $\boldsymbol{\sigma}$ from eq. \ref{sigmaLE},
\begin{equation} 
    \mathcal{G} \ast (\mathbb{C}(\mathbf{x}) : (\boldsymbol{\varepsilon}-\boldsymbol{\varepsilon}^{EIG})) = \boldsymbol{0}.
\end{equation}

The total strain $\boldsymbol{\varepsilon}$ is decomposed into a homogeneous average strain tensor, represented by the macro-scale applied strain $\boldsymbol{E}^{T}$, and a periodic fluctuating micro-scale strain field $\boldsymbol{\Tilde{\varepsilon}}$, which is the unknown. 
\begin{equation}
    \boldsymbol{\varepsilon} = \boldsymbol{\Tilde{\varepsilon}}+\boldsymbol{E}^{T}
\end{equation}
Combining the last two equations results in a linear system in which the strain fluctuation at the voxels are the unknowns. 
\begin{equation}
    \mathcal{G} \ast (\mathbb{C} (\mathbf{x}) : \boldsymbol{\Tilde{\varepsilon}} ) = -\mathcal{G} \ast (\mathbb{C} (\mathbf{x}) : (\boldsymbol{E}^{T} - \boldsymbol{\varepsilon}^{EIG})) 
\end{equation}
This equation can be solved easily by transformation to Fourier space where the operator has a closed expression and convolutions are transformed into products. If a linear discrete operator $\mathcal{G}_C ( \bullet )$ is defined as
\begin{equation}
    \mathcal{G}_C ( \bullet ) = \mathcal{F}^{-1} ( \Hat{\mathcal{G}} :  \mathcal{F} (\mathbb{C} (\mathbf{x}) : ( \bullet )) )
\end{equation}
with $\mathcal{F}$ and $\mathcal{F}^{-1}$ the discrete Fourier transform operations and $\Hat{\mathcal{G}}$ the projection operator in Fourier space, the equation to solve is
\begin{equation}
    \mathcal{G}_C ( \Tilde{\varepsilon} ) = - \mathcal{G}_C ( \boldsymbol{E}^{T} - \boldsymbol{\varepsilon}^{EIG}) ).\label{eq:linear_equation}
\end{equation}
To solve this linear problem efficiently, an iterative solver has to be used. The convergence rate and quality of the solution depend on the phase property contrast, which here is infinite because one of the phases is empty. Following  \cite{LUCARINI2022114223} this problem can be solved efficiently and with minimal noise by using a discrete projection operator (here rotated discrete $\mathcal{G}$-operator \citep{willot2015fourier}) and a minimum residual iteration method as linear solver. 

It is important to highlight that the framework presented is thought to study general geometrical domains in 3-D with heterogeneous microscopic fields. While this may not be regarded as fundamental for a simple cylindrical pillar, it is essential for the modeling of complex geometries, such as nanoporous metals and nanoarchitected meta-materials.

For a set of initialization parameters, our method can be summarized by the algorithm \ref{algoritmo} presented in the Appendix. 

\section{Application to single-crystalline W pillar compression}
\label{sec_resultsGeneral}

Tungsten is a prototypical refractory metal with a high melting point and good corrosion resistance. Its outstanding mechanical properties at high temperatures, together with its low sputtering rate, make this material an ideal candidate for plasma-facing components in fusion energy devices \citep{rieth2013recent}. In addition, tungsten is one of the most studied bcc metals by micropillar compression testing \citep{schneider2009correlation,kim2010tensile,abad2016temperature,srivastava2021influence} and its dislocation mobility laws are well-documented \citep{cereceda2013assessment,cereceda2016unraveling,po2016phenomenological}. These aspects make W an ideal candidate to test the applicability of our framework. 

Single crystalline tungsten pillars were simulated using the framework described in the previous section. The pillar dimensions were varied in the range from 25 nm to 1 $\mu$m at a strain-rate of $10^{-3}$ $s^{-1}$. The crystalline orientation and loading axis were consistent with the [100] direction. 
In all cases, an initial dislocation density of $\rho=5\cdot 10^{12}~m^{-2}$ was taken. Note that such an election renders an initial dislocation count $N_d$ in the range of 1 for the smallest pillars to 8 for the largest ones. The number of potential nucleation sites was taken as $N_{SN} = 128$. The parameters of the surface nucleation model are presented in Table \ref{table:SN1}. The parameters of the physically-based dislocation mobility law are presented in Table \ref{table:dislocmobility}. A list of the slip systems considered can be found in Table \ref{Tab:slipsystems}. 
The models used were discretized in a grid of $128\cdot 128 \cdot 128$ voxels with a kMC normalization factor of $\Delta \varepsilon^*=10^{-3}$. See Supplementary Material file for sensitivity of the results to the discretization.

 In particular, we focus on three important aspects: size effects (Sec.\ref{sec:sizeeffects}), strain-rate effects (Sec.\ref{sec:strain-rateeffects}) and mechanical annealing effects (Sec.\ref{sec:mechanicalannealing}). For validation purposes, the results of the model are later compared with experimental results in Sec. \ref{discussion}. 

\subsection{Size effects}
\label{sec:sizeeffects}
\begin{figure}[H]
    \centering
    \includegraphics[width=1.0\textwidth]{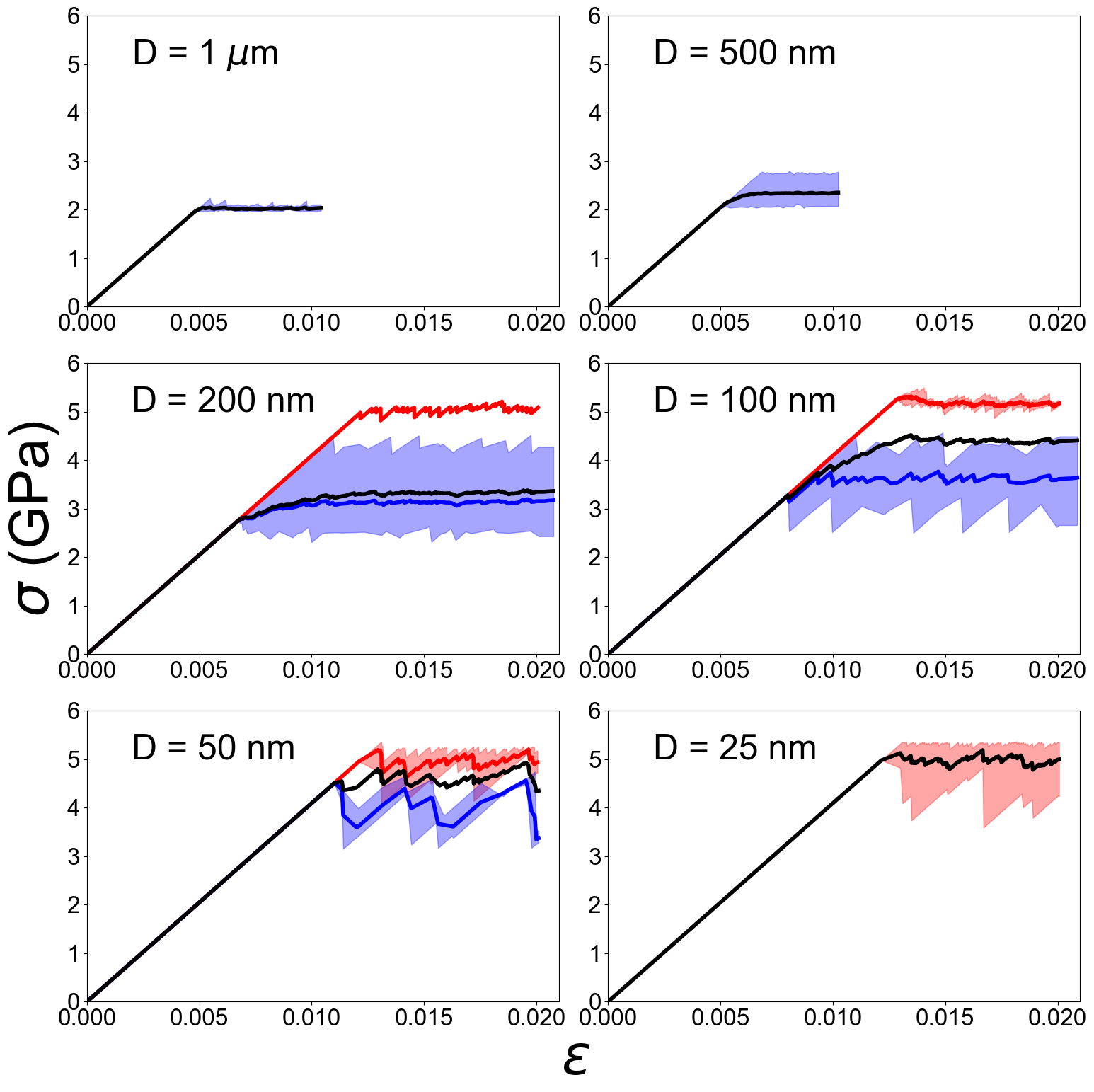}
    \caption{Stress-strain response for W micro/nanopillars under compression. For each of the pillar dimensions probed, ten simulations were
conducted. Shaded regions correspond to the envelope of the mechanical response for all the stress-strain curves, where  blue corresponds to single-arm source dominated cases and red corresponds to surface nucleation dominated cases. Blue and red lines correspond to the average of each subset (SAS-dominated and SN-dominated, respectively), whereas black lines correspond to the average of all the results, irrespectively of their nature.}
    \label{fig:size_effects}
\end{figure}

Figure \ref{fig:size_effects} presents the stress-strain curves for all the simulations performed in this study. For each of the pillar dimensions probed, ten simulations were conducted. 
Blue lines correspond to source truncation (SAS) dominated cases, whereas red lines correspond to surface nucleation-dominated ones. 
The envelope of the mechanical response for all the stress-strain curves is illustrated by the corresponding colored areas. Black lines correspond to the average of the results.

For the largest pillars, with a diameter of  1 $\mu m$, the plastic regime is characterized by a rather smooth output with minimum scatter among the curves. This is a consequence of having several single-arm sources distributed along the sample, allowing the activation of several slip planes in a sequential way. For 500 nm diameter pillars, the plastic regime is again smooth, but now the curves appear more scattered, as judged by the upper and lower limits of the blue-shaded region. For a smaller pillar diameter, the average number of dislocations is close to one. As their pinning position is randomly selected, significant scatter is then translated to the loading curves. The mean flow stress is higher than before, as expected for a decreased pillar diameter. 

For 200 nm diameter pillars, again a significant scatter is seen for the loading curves. Interestingly, one case rendered a significantly high flow stress, associated to surface nucleation of dislocations instead of the activation of the SAS mechanism. This corresponds to the red curve, for which a SAS length of 8 nm was randomly generated. The short length thus renders an activation stress that is higher than that of the SN model. This behavior is further analyzed in the Supplementary Material. 

For 100 nm diameter pillars, now several cases appear both for the activation of SA sources (blue) as well as surface nucleation (red). Both the mean flow stresses for the SAS cases as well as the average (black) increase with respect to the previous pillar diameter. For 50 nm diameter pillars, the surface nucleation mechanism is dominant, with few cases of activation of SA sources at higher stresses compared to previous dimensions. Finally, for 25 nm diameter pillars, surface nucleation was found to be the only operating mechanism. Note that as surface nucleation becomes the dominant mechanism, the curves show smaller scatter. This is consistent with experimental observations \citep{kiener2011source,huang2011new}.


\subsection{Strain-rate effects}
\label{sec:strain-rateeffects}

\begin{figure}[H]
    \centering
    \includegraphics[width=1.0\textwidth]{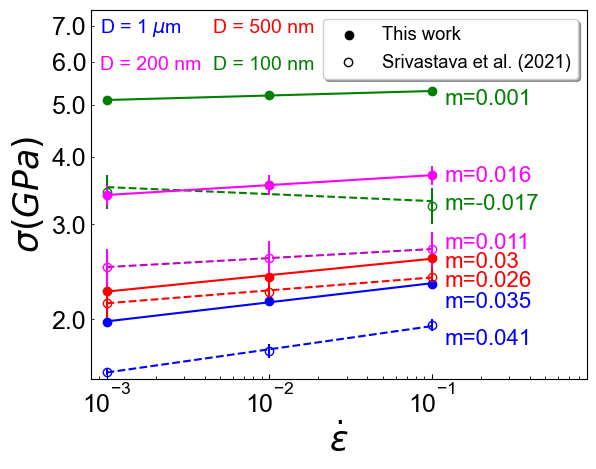}
    \caption{Size-dependent strain-rate effects on the flow stress of W pillars. Empty circles and dashed lines correspond to experimental results by \cite{srivastava2021influence}. For comparison purposes, the colors are the same as the ones used in \cite{srivastava2021influence} and correspond to: blue - 1 $\mu$m; red - 500 nm; magenta - 200 nm; green - 100 nm.}
    \label{fig:strrate_effects}
\end{figure}

In order to evaluate the strain rate sensitivity of the flow strength, W pillar simulations were conducted at different strain rates ($10^{-3}/s,~10^{-2}/s$ and $10^{-1}/s$) and for different pillar diameters (1 $\mu$m, 500 nm, 200 nm and 100 nm). 
Figure \ref{fig:strrate_effects} presents the logarithm of the flow stress as function of the logarithmic applied strain rate obtained by the model for pillars of different diameter and compared to that of \cite{srivastava2021influence} at room temperature.
The log-log flow stress data was then fitted with a power-law function of the strain rate, $\sigma \propto \dot{\epsilon}^m$. The strain rate sensitivity parameter $m$ is then $m=\partial(ln \sigma) / \partial(ln \dot{\epsilon})$.

The resulting curves are approximately linear in the double-logarithm diagram, indicating a strain rate sensitivity $m$ approximately constant in the strain range explored. Comparison with \cite{srivastava2021influence} shows not only similar strain rate sensitivity parameters but also a similar transition trend from positive to null values as the pillar diameter decreases.  

\subsection{Mechanical annealing}
\label{sec:mechanicalannealing}

\begin{figure}[H]
    \centering
    \includegraphics[width=1.0\textwidth]{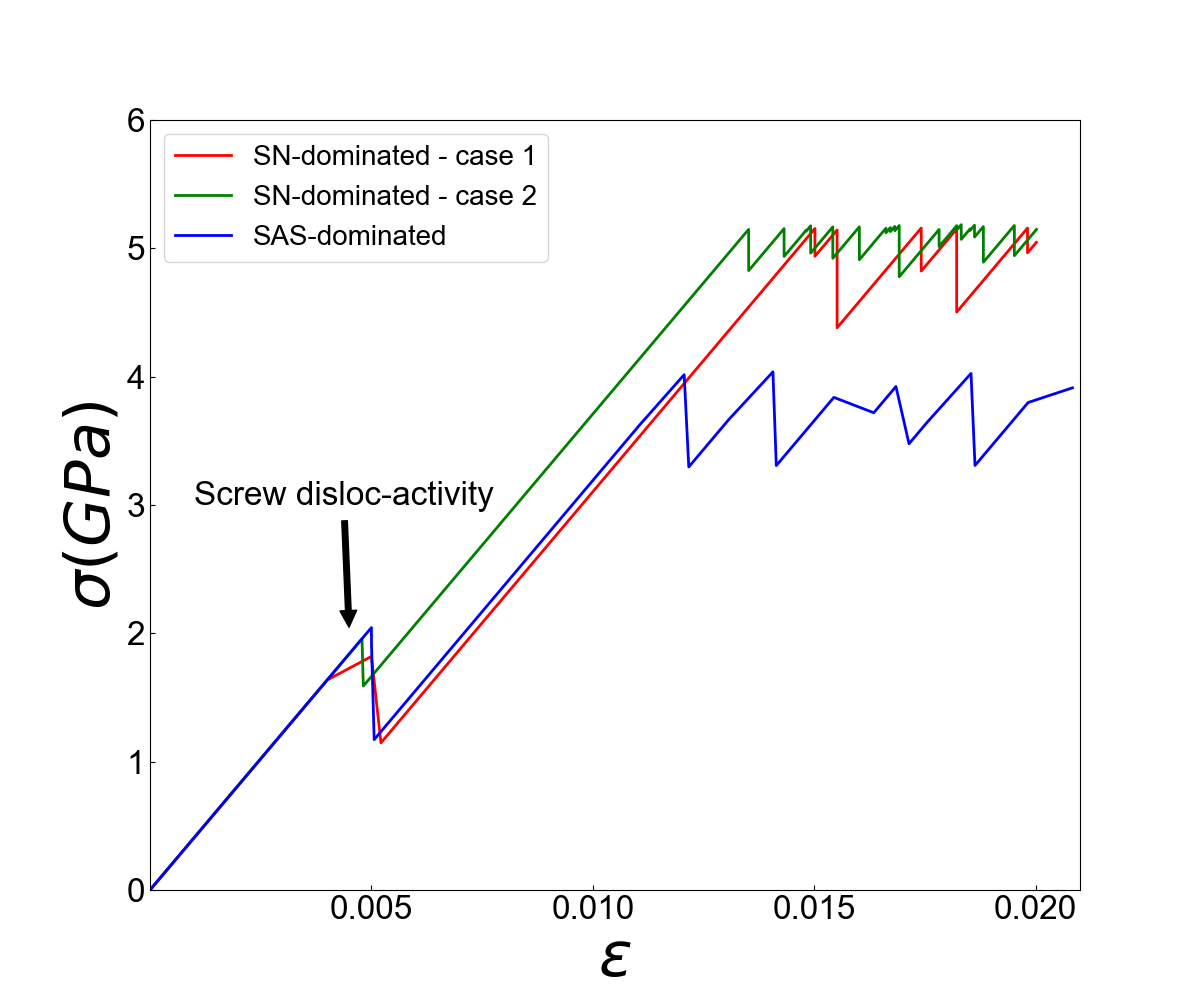}
    \caption{Stress-strain response for 100 nm diameter pillar simulations containing a pure screw dislocation. Red and green curves correspond to pillars without SASs and a screw dislocation located in the middle of the cross section (red - case 1) and halfway between the center of the pillar and its surface (green - case 2). Blue curve corresponds to a pillar with a sufficiently long SAS and a screw dislocation located in the middle of the cross section.}
    \label{fig:mech_anneal}
\end{figure}

In addition to single-arm sources, pre-existing pure screw dislocations can also be considered, as presented in the methods section. These are valuable for capturing mechanical annealing effects \citep{shan2008mechanical,huang2011new,shan2012situ} associated with dislocation starvation \citep{greer2006nanoscale}. 
Figure \ref{fig:mech_anneal} presents three distinctive stress-strain curves corresponding to 100 nm diameter pillar simulations where one pre-existing screw dislocation has been considered. The red curve corresponds to a pillar without SASs and a screw segment located in the middle of the cross-section (SN-dominated - case 1), while the green curve corresponds to a pillar without SASs and a screw segment located halfway between the center of the pillar and its surface (SN-dominated - case 2). In both cases, after the operation of the screw dislocation, the source has exhausted. In consequence, the simulated test continues elastically until activation of the SN mode. A similar scenario would have been found for a case in which a pre-existing screw dislocation habits a pillar with a SAS of sufficiently small length. In our implementation, the position of the screw segment is randomly chosen in the pillar cross section considering a uniform distribution. Specific knowledge of the material under consideration would allow to modify such distribution accordingly and without invalidating our methodology. Differences in load drops for the activation of screw segments are discussed in Sec. \ref{discussion}.
In contrast, the blue curve corresponds to a pillar with a screw dislocation in the middle of the cross section and a sufficiently long SA source. Upon loading, and after an initial elastic stage, the screw dislocation is activated -recall that the critical resolved shear stress is lower than that of SASs -. After exhaustion of the screw source, the mechanical response proceeds elastically until reaching a stress corresponding to the CRSS of the SA source, which gets activated several times for the rest of the simulation. 

\section{Discussion}
\label{discussion}

\paragraph{\bf{Comparison with other methods}}

Several aspects of our proposed framework are worth discussing. 
The framework proposed aims at modelling the full elastic-plastic response with 3D resolution of microfields. Its particular application to nano-pillars and submicron-pillars allows to obtain the size effect in the yield like other simple models \citep{parthasarathy2007contribution,jennings2013modeling}, but here the microfields and post yield behavior are also recovered, with the aided benefit of obtaining the 3-D deformation pattern (see below).
Our model can therefore be placed in between DDD approaches and continuum descriptions, such as crystal plasticity. Regarding CP, continuum microfields are used but deformation is localized here. In this aspect, our method shares a similarity with the concept of thin discrete slip bands used by \cite{wijnen2021discrete} but the nature of our model is stochastic and based on the displacement discontinuity created by slip events. 
Like in DDD approaches \citep{el2009role,srivastava2013dislocation,cui2014theoretical,ryu2013plasticity,ryu2015stochastic,ryu2020intrinsic}, important aspects of the dislocation mobility laws and underlying physics are considered. The selection of the plastic events does not take place using the weakest-link principle, like in \citep{el2009role,wijnen2021discrete}, but a Monte Carlo selection process over the displacement rates calculated using an approach similar to Orowan's equation. 

However, unlike DDD approaches, dislocations are not modeled explicitly, but their effect is introduced by means of the concept of Eshelby inclusions, such that the effect of dislocation gliding is introduced through an equivalent eigenstrain. The total strain is thus the superposition of two fields, ($\boldsymbol{\varepsilon}$ and the inelastic strain obtained as a sum of eigenstrains due to slip events $\boldsymbol{\varepsilon}^{EIG}$). Therefore, the calculation of the strain and stress fields becomes linear, in contrast to models based on localized crystal plasticity as \citep{wijnen2021discrete}. This aspect, together with an inherently fast FFT solver, results in an efficient and computationally cheap stochastic framework capable of simulating experiments with 3D spatial resolution at strain rates lower than $10^{-3}/s$ on nano and submicron samples within a few hours on an off-the-shelf workstation. The present implementation is based on small strains since we are focusing on the initial flow of the pillars, but can be extended to finite strains by assuming a non-linear elasticity energy and solving the FFT problem in finite strains \citep{zeman2017finite}. Other aspects for further extension of the model include non-Schmid effects. This could be done following steps already taken with recent advances on the development of a physically-informed continuum crystal plasticity model for tantalum \citep{lee2023deformation}. 
The framework would also allow for the incorporation of twinning as a complementary deformation mechanism. Discrete dislocation plasticity models have proven useful for the understanding of dislocation slip-mediated twinning mechanisms \citep{wang2022evolution} and could be used as input for the extension of the framework presented here.

\begin{figure}[H]
    \centering
    \includegraphics[width=0.95\textwidth]{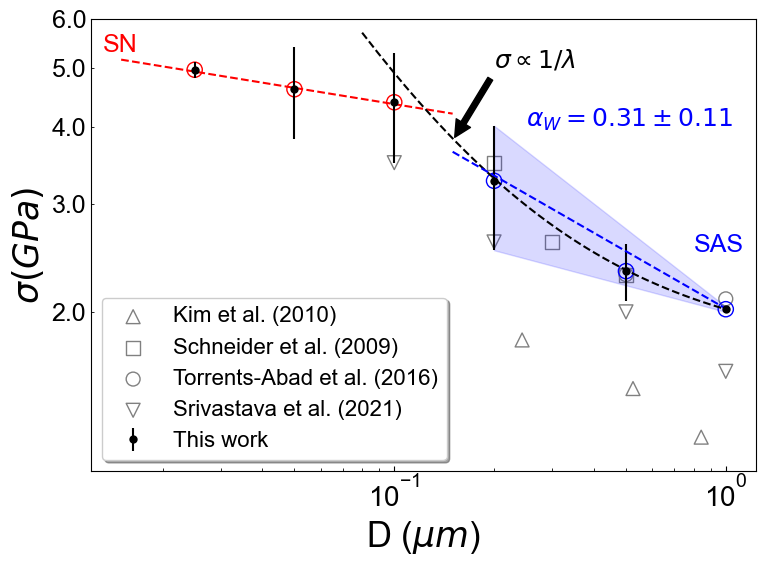}
    \caption{Log-log size-dependence of the flow stress compared with experimental literature. Black dashed line indicates $1/\lambda$ dependence, in agreement with a SAS dominated regime. Blue dashed line indicates the estimated power law exponent, while light-blue shadow indicates $\pm$ values ($\alpha_W \approx 0.31 \pm 0.11$). Red dashed line indicates a surface nucleation (SN) trend. Its intersection indicates a cross-over transition size. Vertical bars correspond to standard deviation.}
    \label{fig:summary}
\end{figure}

\paragraph{\bf{Flow stress size dependence}} The proposed framework allows to obtain the statistical distribution of the mechanical response of a pillar as a function of its size, as shown in Fig \ref{fig:size_effects}. To analyze these results, in Figure \ref{fig:summary} the flow stress for different pillar sizes obtained with our model is compared to selected experimental references \citep{schneider2009correlation,abad2016temperature,srivastava2021influence,kim2010tensile} in a double logarithm representation. Flow stress values presented correspond to the average stress in the plastic regime of the curves presented in Fig. \ref{fig:size_effects}, while vertical bars correspond to standard deviation of all the results presented in Fig. \ref{fig:size_effects}, capturing the dispersion of results produced by the stochastic selection of the SAS length.
\cite{schneider2009correlation} performed load-controlled compression tests on [100]-oriented W pillars spanning from the nm scale (200 nm) to the micrometer scale (6 $\mu m$). Later on, the same group produced an experimental study probing temperature effects on W pillars with similar orientation \citep{abad2016temperature}. 
Our model successfully captures these experimental measurements as well as an overall $\sigma \propto 1/\lambda$ dependence of the yield strength in the 200 nm - 1 $\mu m$ range (black dashed line). \cite{srivastava2021influence} performed displacement-controlled compression tests on [100]-oriented W nanopillars in the range of 100 nm - 1 $\mu m$. Their results are systematically lower than those reported in \cite{schneider2009correlation,abad2016temperature} yet following a similar yield strength trend increase with decreasing size ($\sigma \propto 1/\lambda$). \cite{kim2010tensile} performed displacement-controlled compression tests on [100]-oriented W nanopillars in the range of 200 - 900 nm. Their results represent a lower bound for the experimental data set chosen. Yet their values again follow similar trends ($\sigma \propto 1/\lambda$).
To the best of the authors knowledge, there are no reports on [100]-oriented W pillars with diameters below 100 nm, probably due to the associated experimental difficulties. At the limit of a 100 nm, we predict higher yield stresses than those reported in \cite{srivastava2021influence}, yet our estimations agree with their measurements if we take into account the standard deviation.

For diameters below 200 nm, the exponential dependence $\sigma \propto 1/\lambda$ yields to a surface nucleation-dominated trend (red dashed line), indicating a crossover between predictions from the SAS and SN models. The critical sample size observed is approximately 120 nm. Below this threshold, the flow stress determined by the SN model is lower than that by the SAS model, highlighting the comparative difficulty of activating a single-arm source over nucleating a dislocation from a free surface, emphasizing a preference for surface nucleation.
In contrast, when dealing with samples that exceed this critical size, plasticity is typically ruled by the activation of pre-existing dislocation sources, such as those represented by the SAS mechanism. A handful of previous studies had focused on the critical size for bcc and fcc metals using both experimental and computational techniques. For fcc materials, \cite{hu2019predicting} used DDD simulations to predict the the flow stress and dominant yielding mechanisms in Cu nanopillars ($100~nm < D < 800~nm$), reporting on a cross-over from SAS activation to SN operation as D reaches $\approx 110 nm$. Similarly, \cite{shan2008mechanical} utilized transmission electron microscopy (TEM) to investigate the behavior of pre-existing dislocations in Ni pillars of different diameters. The authors observed that in the case of a 160 nm diameter pillar, the pre-existing dislocations gradually exited the pillar, followed by the emergence of new dislocations, which can be attributed to the phenomenon known as SN. Conversely, in the case of pillars with a diameter of 290 nm, dislocations persisted even after undergoing compression, which can be tentatively associated with SAS. For bcc metals, experiments conducted on bcc Mo pillars suggest a critical size of around 200 nm \citep{huang2011new,shan2012situ}. 
We conclude that our critical size prediction for [100]-oriented W pillars is in agreement with previous observations for other metals at these scales.


The results can be interpreted through the well-known power law relation, namely
\begin{equation}
    \label{power_law_alpha}
    \sigma_y = \sigma_0 + k_1 D^{-\alpha_W}
\end{equation}
where $\sigma_0$ is the strength of the bulk material, $k_1$ is a constant and $\alpha_W$ is the power law exponent. In practice, $\sigma_0$ is usually neglected \citep{brinckmann2008fundamental,abad2016temperature,srivastava2021influence} as the second term in eq. \ref{power_law_alpha} dominates at these small scales. As a reference, \cite{srivastava2021influence} report on exponent of 0.319 for their $10^{-3}~s^{-1}$ experiments, close to the $\alpha_W=0.33$ value obtained by \cite{abad2016temperature} for the same strain rate. In turn, \cite{kim2010tensile} informed a power law exponent in the range of $0.33 < \alpha_W < 0.55$. On a lower end, \cite{schneider2009correlation} report on $\alpha_1 = 0.21$. By linear fitting eq. \ref{power_law_alpha}
on the SAS-dominated regime ($D > 120 ~nm$), we obtain an exponent $\alpha_W \approx 0.31 \pm 0.11$, in agreement with most of the experimental evidence. This further supports the ability of the single-arm source model to understand the power-law relation on the strength vs size dependence of micropillars \citep{lee2012size}.  
The red dashed line, connecting the average yield stress values obtained for the smaller pillars, presents a slope that seems to be an indication of a possibly weaker size dependence.

The underlying cause for the values of $\alpha_W$ are worth a discussion. \cite{brinckmann2008fundamental} used the screw dislocation cross-slip mechanism concept, derived from high strain-rate MD and DD simulations \citep{weinberger2008surface,greer2008comparing}, to provide an explanation for their bcc Mo pillar compression test results. They argued that the relatively low power law exponent of bcc metals ($\alpha_{Mo}=0.45$) could be attributed to such effects. Interestingly, \cite{kim2010tensile} performed pillar compression tests on W, reporting on strikingly similar power-law exponents ($\alpha_{W}=0.44 \pm 0.11$). One could then assume that the same mechanism takes place in bcc W. Yet, our power-law exponent estimations are on the lower bound of this regime, while our model does not incorporate a single characteristic of the cross-slip mechanism presented in \citep{weinberger2008surface}. Therefore, our results suggest that screw dislocation mobility alone could indeed provide adequate power-law exponents provided the underlying physics of the dislocation mobility law are fully incorporated ( eq. \ref{v_alfa}). 

It is worth highlighting that previous experimental observations on bcc Mo suggest that size effect itself has a strong size effect \citep{huang2011new,shan2012situ}. In their in-situ TEM bcc Mo studies, \cite{huang2011new} found a strong increase in the power law exponent as the pillar diameter decreases beyond 200 nm. 
In contrast, we predict a break down in the power-law exponent as the W pillar diameter decreases below 200 nm. This difference in behavior could be attributed to different activation energies between Mo and W. Small variations in activation energy can strongly influence yield stress at the nanoscale. 
A break down in exponent can also be inferred from dislocation dynamics simulations on Cu \citep{hu2019predicting}. 
We attribute the break down in power law exponent to the change of mechanism, from SAS-dominated to surface nucleation dominated, for which a weaker stress dependence on diameter is expected \citep{zhu2008temperature,huang2015flow,hu2019predicting}.  Certainly, this alternative explanation of the results leaves space for future research and informed discussion.

\paragraph{\bf{Strain-rate sensitivity}} Our results show a size dependence of the strain rate sensitivity parameter, which changes from positive $m$ to nearly null values (Figure \ref{fig:strrate_effects}). \cite{huang2015flow} and \cite{srivastava2021influence} observed the same positive to null trends for their bcc Fe and W pillar experiments, respectively. 
The strain rate sensitivity exponent $m$ depends on the activation volume $V_a$ and on the resolved stress on the active plane $\tau$ through \citep{kocks1975thermodynamics}:
\begin{equation}
    \label{m_exponent}
    m = \frac{k_B T}{\tau V_a}.
\end{equation}

In the previous equation, both the critical shear $\tau$ and the activation volume $V_a$ can be size dependent, and the origin of size dependency of $m$ might have both contributions. Regarding the effect of activation volume, \cite{huang2015flow} rationalized that in Fe this might be the main contribution to size effect in $m$. This size dependent $m$ in Fe arises from a competition between a size-independent friction term -that rules for large pillar diameters (D $>$ 500 nm)- and a size-dependent term -that rules for smaller pillar diameters-, with larger activation volumes for increasingly smaller pillars (100 nm $<$ D $<$ 500 nm). They also argue that as the pillar diameter further decreases (D $<$ 100 nm), the activation volume becomes again smaller. 
Still, Fe is known to have an activation volume that may display larger variations with stress compared to other bcc metals with higher melting point \citep{christian1964low}. Tungsten, in constrast, and particularly for low homologous temperatures, displays a very low activation volume ($V_a \leq 10b^3$) \citep{kiener2019rate}, consistent with the dominance of the kink mechanism at low temperatures. Our calculations of the activation volume (see Suppl. Mat.) render that for the kink-pair activated mechanism, the stress-dependent activation volume is in the range of 4 $b^3$ $<$ $V_a^{kink}$ $<$ 11 $b^3$ for the stress range of the SAS-dominated regime, whereas for the surface nucleation mechanism is in the range of 2.5 $b^3$ $<$ $V_a^{SN}$ $<$ 4.5 $b^3$ for the stress range of the SN-dominated regime. 

Therefore, considering the relatively low variations in the activation volume $V_a$ in W, changes in the strain rate sensitivity exponent $m$ can be understood based on the size-dependence of the stress reached on the pillar due to the size dependent strength of SAS, as well as on the stress-dependence of the dislocation mobility law. Similar conclusions on the importance of $\tau$ were reached by Srivastava et al. \citep{srivastava2021influence} for their strain-rate sensitivity studies on W pillars.

Future investigations using our proposed framework could also focus on temperature effects, as a recently developed CP model using similar dislocation mobility laws has proven useful in exploring the temperature dependence of  deformation localization in irradiated tungsten 
\citep{li2021temperature}.

\paragraph{\bf{Mechanical annealing}} Earlier MD simulation studies predict that such behavior cannot take place in bcc pillars. Such simulations suggested that the combined effects of image forces, together with the dislocation core structure, would facilitate dislocation multiplication by a surface-assisted cross-slip mechanism \citep{weinberger2008surface}. SEM observations of multiple slip traces on compressed bcc pillars seem to support this mechanism \citep{schneider2009correlation}. On the other hand, in-situ TEM compression of Mo nanopillars \citep{huang2011new,shan2012situ} and  Fe–3\% Si pillars \citep{zhang2012dislocation} show clear evidence that mechanical annealing by dislocation starvation can indeed take place in bcc metals. 
Our proposed computational framework takes into account such scenarios. Interestingly, load drops due to mechanical annealing are, in general, less marked than in our Figure \ref{fig:mech_anneal}, as shown in \citep{shan2008mechanical}. Other works do report stress drops of the order of 0.5 - 1.0 GPa \citep{huang2011new,zhang2012dislocation}. Both situations can be taken into account in our method by an adequate selection of the statistical distribution of pre-existing screw segments.

\begin{figure}[H]
    \centering
    \includegraphics[width=1.0\textwidth]{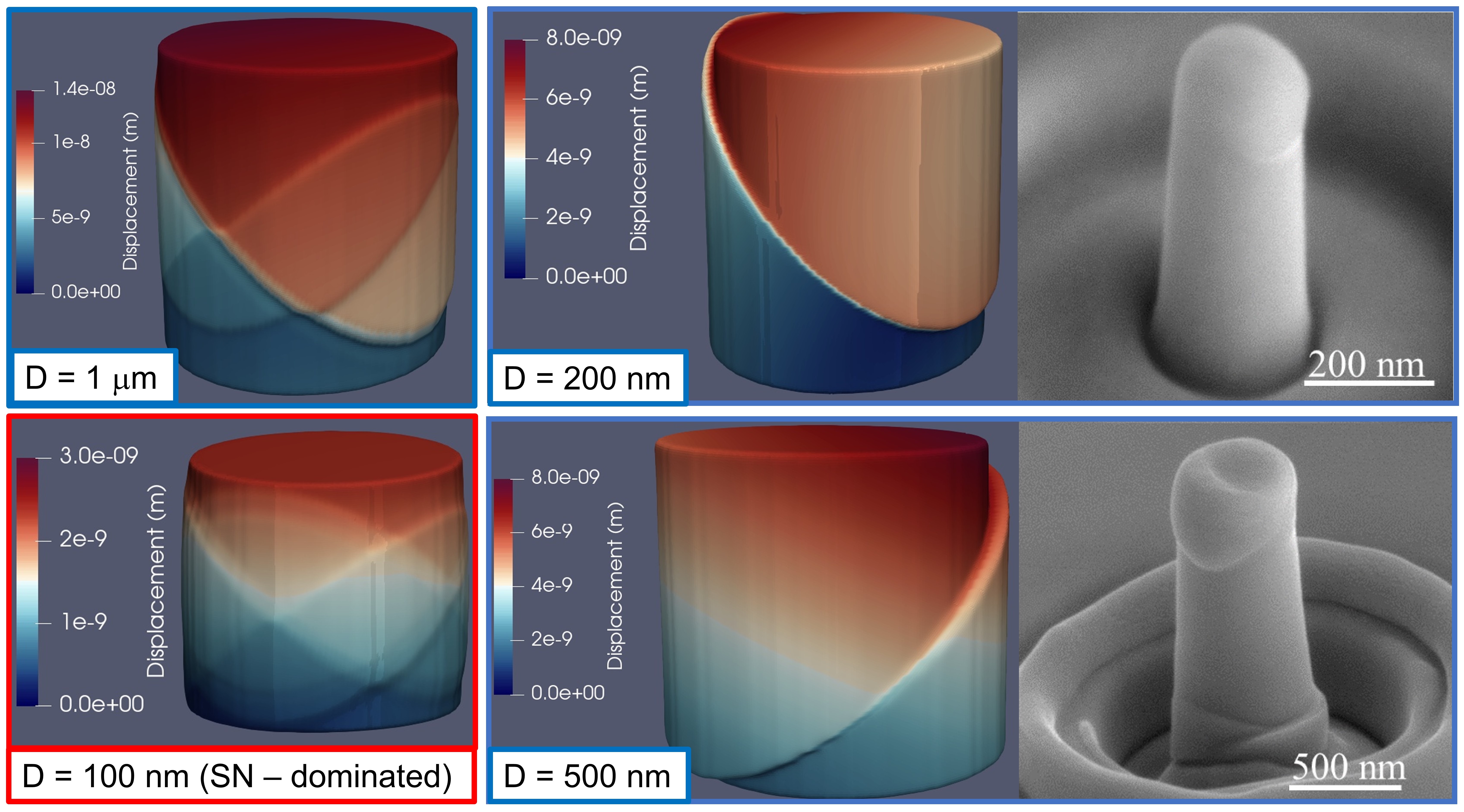}
    \caption{Residual deformation patterns for different pillar diameters explored, together with SEM observations by \cite{srivastava2021influence}. Reprinted with permission from Elsevier. Frame colors correspond to dominating deformation mechanism (Blue - Single-Arm source, Red - Surface nucleation).}
    \label{fig:deformation}
\end{figure}

\paragraph{\bf{Deformation patterns and localization}} Figure \ref{fig:deformation} presents the typical deformation patterns obtained in our simulations, together with a comparison with a selected reference \citep{srivastava2021influence}. For our largest pillar diameter ($D = 1 \mu m$), the deformation pattern reveals the activation of several SASs on a variety of slip systems (see Supplementary Material movie), in agreement with scanning electron microscopy (SEM) analysis of W micron-sized pillars \citep{schneider2009correlation,abad2016temperature}. As the pillar diameter decreases to the submicron regime ($D = 500~nm$ and $D = 200 ~nm$) and for the dislocation density chosen, only one SAS is present. Upon activation, significant deformation takes place and concentrates on a single slip band, in agreement with SEM micrographs of W pillars with comparable diameters \citep{srivastava2021influence}. Further decrease of pillar diameter into the nanometer regime ($D=100~nm$) triggers surface nucleation instead of SAS activation. The former takes place on multiple slip systems.  

As explained in Sec. \ref{kMC_details}, our model favors the activation of weaker sources. Localization of plastic activity in one or a few number of planes depends on the number of potential slip sites (e.g. number of SAS) and the statistical distribution of the slip resistance on that planes (linked with the length of each SAS). Our 1 $\mu m$ diameter pillars with $\rho=5\cdot 10^{12}~m^{-2}$ have multiple SASs of variable length. They show a tendency for a rather delocalized (homogeneous) distribution of deformation. In contrast, 200 nm - 500 nm diameter pillars with similar dislocation density have 1 - 2 SASs on average. In such cases, localization is favored (Figure \ref{fig:deformation}). \cite{cui2018size} presented a 2D Monte-Carlo model of DD source activation coupled with crossslip channel widening to reproduce and physically explain the transition in the mechanism of plastic flow localization in irradiated materials from irradiation-controlled to dislocation source-controlled. They showed that as the size decreases, the spatial correlation of plastic deformation decreases due to weaker dislocation interactions and less frequent cross-slip, thus producing thinner dislocation channels. For a low irradiation damage, they show that as the diameter decreases from 1.5 $\mu m$ to 300 nm, deformation transitions from rather homogeneous to highly localized. Our results broadly agree with this picture.

\begin{figure}[H]
    \centering
    \includegraphics[width=0.95\textwidth]{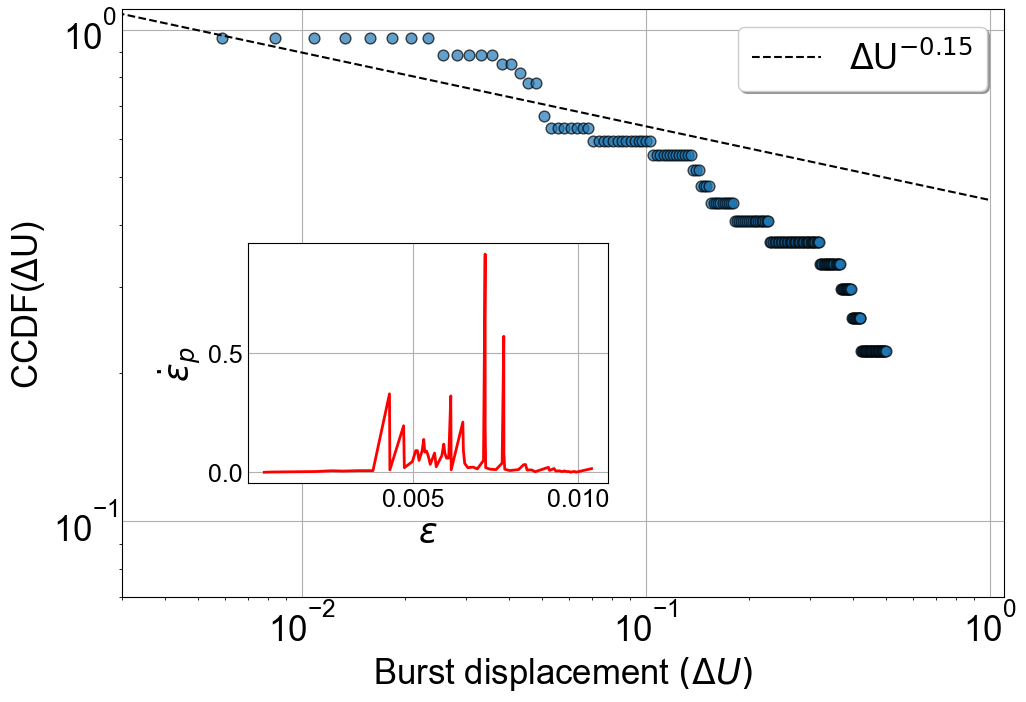}
    \caption{Complementary cumulative distribution function (CCDF) of burst displacement for one example of our 1 $\mu$m diameter pillars. Inset, corresponding evolution of plastic strain rate showing typical quasiperiodic strain bursts, consistent with displacement-controlled compression conditions. Note the breakdown from power law scaling (dashed line).}
    \label{fig:bursts}
\end{figure}

\paragraph{\bf{Statistical analysis of burst displacement}}
In order to analyze the statistics of the plastic bursts obtained in the simulations, the complementary cumulative distribution function (CCDF) of the burst displacement magnitude $\Delta U$ is obtained.  The CDDF is computed using the method presented by \cite{cui2016controlling} and its result for  a 1 $\mu$m diameter pillar is represented in Figure \ref{fig:bursts}.
The Figure suggest that, under the displacement-controlled conditions (strain control) explored here, the system does not exhibit power-law scaling, in agreement with \cite{cui2016controlling}. In addition, the data spans less than two orders of magnitude. In contrast, typical load-displacement controlled compression experiments usually span several orders of magnitude, with bursts often exceeding 10 nm \citep{alcala2020statistics,srivastava2021influence}. Detailed tracking of the plastic strain rate (Inset of Figure \ref{fig:bursts}) shows that the dynamical behavior is quasiperiodic, in agreement with displacement-controlled simulations \citep{cui2016controlling} and experiments \citep{papanikolaou2012quasi}. In contrast, experiments under load-controlled conditions lead to a completely different scenario, where plasticity does not lead to marked stress drops, allowing to trigger the operation of several SAS simultaneously and thus exhibiting power law scaling. As a result, avalanche events can develop, leading to a highly correlated dynamical response, consistent with the concept of self-organized criticality \citep{bak1988self}. The framework presented here, inherently implies the activation of one source at a time, leading to a quasiperiodic response that it is consistent with displacement-control conditions.

\paragraph{\bf{On the incorporation of strain hardening effects}}  In all the cases presented in Figure \ref{fig:size_effects}, the flow stress keeps constant, in agreement with the assumptions introduced in Sec. \ref{sec:methods}. This is consistent with an scenario in which the slip planes swept by the SAS exhibit a scarcity of dislocations in intersecting planes, and where the pinning point of the SAS is stable under the prevailing stress levels. Consequently, the activated SAS does not encounter other defects, hence preventing the formation of new junctions or the reduction of the source length. In addition, experiments on W single crystalline pillars \citep{schneider2009correlation} do not show hardening for the relatively low strains explored in Figure \ref{fig:size_effects}. 

For larger samples, our method could be extended to incorporate hardening effects by including a Taylor hardening-like term in eq. \ref{SAS_CRSS} and a dislocation density evolution law \citep{cui2014theoretical}. The latter should incorporate size-dependent terms and bulk-like terms. Size-dependent terms include the generation of dislocations due to the operation of existing sources ($\propto 1/\lambda$) and the escape of dislocations reaching a free surface ($\propto - 1/D$) \citep{cui2014theoretical}. Bulk-like terms include dislocation multiplication due to forest dislocations ($\propto \sqrt{\rho}$) and the annihilation of closely spaced dislocations of opposite signs ($\propto - \rho$) \citep{devincre2008dislocation}. If self and pair interactions must be taken into account (e.g. for micron scale pillars above 1 $\mu$m), the corresponding hardening coefficient matrix can also be incorporated \citep{yu2021stochastic,lee2023deformation}. It must be emphasized that for submicron pillars, simple calculations show that the contributions of the bulk-like terms are nearly an order of magnitude smaller than the size-dependent terms \citep{cui2014theoretical}. Thus, by neglecting the former and provided the dislocation generation rate is balanced with the escape rate from free surface which, together with a relatively low dislocation density, the results presented in Sec.\ref{sec:sizeeffects} would be reproduced.

To wrap up, the results presented here focus solely on tungsten and a relatively simple geometry. Yet, these are not restrictions. For instance, FeCrAl alloys and refractory multi-principal element alloys (RMPEAs) are also promising candidate materials for use under extreme conditions in the nuclear industry. FeCrAl alloys have already been probed using discrete dislocation dynamics \citep{pachaury2023plasticity} as well as with crystal plasticity approaches using Arrhenius-type rate equations \citep{gong2023bridging}. With respect to RMPEAs, kink-migration of screw dislocations have been shown to play a dominant role in their deformation \citep{balbus2024orientation} and it is expected that developments in dislocation mobility laws \citep{shen2021mobility} would allow for systematic investigations on these promising alloys using upper scale methods. Our proposed framework offers flexibility to explore the micropillar compression behavior of
such alloys by the incorporation of specific dislocation mobility laws. To further extend its potential, the use of a 3D FFT solver offers opportunities for the exploration of a variety of complex geometries, including nanoporous metals and nanoarchitected meta-materials.

\section{Conclusions}

A novel approach is presented to simulate the deformation of submicron specimens, accounting for the discrete events produced by the slip of internal dislocation and surface nucleated ones in a stochastic manner. The framework considers the slip events as eigenstrain fields that produce a  displacement jump across a slip plane and whose activation is driven by a Monte Carlo (MC) method. Physically-based laws are incorporated to account for activation probabilities, dislocation mobility and surface nucleation. 

Two factors of stochasticity are taken into account: a random selection of the position of defects and a random selection of plastic events after a MC process on a sampling array of possible displacement rates, computed using a similar concept as in Orowan's equation. 
Implementation on a fast FFT solver, results on an efficient, computationally-cheap algorithm which allows to simulate accurately the deformation of a specimen in a wide range of strain rates and sizes in a fraction of the time needed using techniques as MD or DDD.

The framework developed has been applied to the compression of tungsten (W) on [100]-oriented pillars in the 25 nm to 1 $\mu m$ diameter range. 
We find an adequate agreement between simulations and experimental data. 
The study a size-dependence of flow stress, characterized by distinct power-law exponents for regimes dominated by source truncation (D $>$ 120 nm, $\alpha_W \approx 0.31 \pm 0.11$) and a break down of the power law scaling as surface surface nucleation becomes dominant (D $<$ 120 nm). Pre-existing dislocations naturally compete with surface nucleation of new dislocations. The former are favored for pillars above 120 nm diameter, whereas the latter is favored below such critical size. In addition, strain-rate sensitivity effects are adequately captured, including a size-dependence of the strain-rate sensitivity exponent. 
This is attributed not only to changes in activation volume and source strength but also to the stress-dependent dislocation mobility law, whose form considers both the kink-pair regime, dominating at the relatively low stresses found in submicron pillars, as well as the phonon drag regime, that rules for very high stresses found in pillars of 200 nm and below. The framework inherently implies the activation of one source at a time, leading to a quasiperiodic response that it is consistent with displacement-control conditions.

Finally, the framework presented here opens the possibility to provide valuable analysis and interpretations of other small-scale testing techniques, such as microtensile testing and microbending. Future developments of dislocation mobility laws for FeCrAl alloys and multi-principal element alloys would allow for a systematic investigation of these relevant metals, while the possibility to include other micro and nanoscale geometries further extends the potential applications map to include nanolattices and nanoporous materials, among other nanostructures of interest. 

\section{Acknowledgements}
This project has received funding from the European Union’s Horizon Europe research and innovation programme under the Marie Sklodowska-Curie grant agreement no. 101062254.
Funded by the European Union. Views and opinions expressed are however those of the author(s) only
and do not necessarily reflect those of the European Union. Neither
the European Union nor the granting authority can be held responsible for them.

\appendix

\section{Appendix Section}
\label{sec:sample:appendix}
\begin{algorithm}[H]
\caption{kMC discrete slip approach with eigenstrains}\label{euclid}
\label{algoritmo}
\begin{algorithmic}[1]
\Procedure{Initialization}{}
 \begin{itemize}
    \item Define pillar geometry, $\mathbb{C}(\mathbf{x})$, in the discrete domain. 
    \item Set positions, eigenstrain fields and strength of each plastic region $i$
    \item Set loading conditions $\mathbf{E}^T(t)$,$\Delta \varepsilon^*$.
\end{itemize}

\EndProcedure

\Procedure{Solution}{}

\While{$t<t_{TOT}$}:

\State Get rnd no. $\xi_1$, $\xi_2$ $\in$ (0,1]
\State Solve the equilibrium (Eq. \ref{eq:linear_equation}) using FFT solver to obtain $\boldsymbol{\sigma}$ (Eq. \ref{sigmaLE}) , $|\tau^i|$ (Eq. \ref{tau_i})
\For{each region} 
\If {$|\tau_{i}| > \tau_{CRSS}$}
\State Compute: $P$ (SN sites only), $v^{i}$ (Eq. \ref{v_alfa}) , $t^{i}$ (Eq. \ref{t_i}) and $\dot{u}^i_p$ (Eq. \ref{eq:u_p2}). 
\Else 
\State $v^{i}=0$ 
\EndIf
\EndFor

\State Compute accumulated rates $r_j$, $r_{tot}$
\State $\delta t^{n+1}=-\frac{log \xi_2~\Delta \varepsilon^* }{r_{tot}}$

\If {$\xi_1<\frac{r_0}{r_t}$}
\State $\delta t^{n+1} = dt^0$
\State next event is elastic: $\boldsymbol{\varepsilon}^{EIG~(i)}=0$
\Else
\State next event is plastic
\State Compute eigenstrain: $\boldsymbol{\varepsilon}^{EIG~(i)}$ (Eq. \ref{eigenstrain2})
\State 
\EndIf
\State Update total eigenstrain field $\boldsymbol{\varepsilon}^{EIG}$ (Eq. \ref{sumEig})

\EndWhile
\EndProcedure
\end{algorithmic}
\end{algorithm}

The slip systems considered in this study are presented in Table \ref{Tab:slipsystems}.
\begin{table}[H]
\begin{tabular}{cccc}
\hline
$i$ & Slip system              & $s^i$            & $n^i$      \\ \hline
1     & $[1\bar{1}1](011)$       & $[1\bar{1}1]$       & $[011]$       \\
2     & $[\bar{1}\bar{1}1](011)$ & $[\bar{1}\bar{1}1]$ & $[011]$       \\
3     & $[111](0\bar{1}1)$       & $[111]$             & $[0\bar{1}1]$ \\
4     & $[\bar{1}11](0\bar{1}1)$ & $[\bar{1}11]$       & $[0\bar{1}1]$ \\
5     & $[\bar{1}11](101)$       & $[\bar{1}11]$       & $[101]$       \\
6     & $[\bar{1}\bar{1}1](101)$ & $[\bar{1}\bar{1}1]$ & $[101]$       \\
7     & $[111](\bar{1}01)$       & $[111]$             & $[\bar{1}01]$ \\
8     & $[1\bar{1}1](\bar{1}01)$ & $[1\bar{1}1]$       & $[\bar{1}01]$ \\
9     & $[\bar{1}11](110)$       & $[\bar{1}11]$       & $[110]$       \\
10    & $[\bar{1}1\bar{1}](110)$ & $[\bar{1}1\bar{1}]$ & $[110]$       \\
11    & $[111](\bar{1}10)$       & $[111]$             & $[\bar{1}10]$ \\
12    & $[11\bar{1}](\bar{1}10)$ & $[11\bar{1}]$       & $[\bar{1}10]$ \\ \hline
\end{tabular}
\caption{Slip systems considered in this study}
\label{Tab:slipsystems}
\end{table}

 \bibliographystyle{elsarticle-harv} 
 \bibliography{cas-refs}





\end{document}